\documentclass[aps,prx,twocolumn,showpacs,preprintnumbers,longbibliography]{revtex4-1}
\usepackage{amsmath}
\usepackage{bm}
\usepackage{epsfig}
\usepackage{color}
\usepackage{comment}
\usepackage{float}
\usepackage{bm}
\usepackage{graphicx}
\renewcommand{\phi}{\varphi}

\newcommand{\eps}{\epsilon}

\newlength{\onefig}
\newlength{\twofig}

\begin{document}

\title{Models and algorithms for the next generation of
glass transition studies}

\author{Andrea Ninarello}

\author{Ludovic Berthier}

\author{Daniele Coslovich}

\affiliation{Laboratoire Charles Coulomb, 
UMR 5221 CNRS-Universit\'e de Montpellier, Montpellier, France}

\date{\today}

\begin{abstract}
Successful computer studies of glass-forming materials need to overcome
both the natural tendency to structural ordering and the 
dramatic increase of relaxation times at low temperatures. We present 
a comprehensive analysis of eleven glass-forming models
to demonstrate that both challenges can be efficiently tackled using 
carefully designed models of size polydisperse supercooled liquids 
together with an efficient Monte Carlo algorithm where
translational particle displacements are complemented by swaps 
of particle pairs. We study a broad range of size polydispersities,
using both discrete and continuous mixtures, and we
systematically investigate the role of particle softness, attractivity and
non-additivity of the interactions.
Each system is characterized 
by its robustness against structural ordering and by the 
efficiency of the swap Monte Carlo algorithm. 
We show that the combined optimisation of the potential's softness, 
polydispersity and non-additivity 
leads to novel computer models with excellent 
glass-forming ability. 
For such models, we achieve over ten orders of 
magnitude gain in the equilibration 
timescale using the swap Monte Carlo algorithm, {thus 
paving the way to computational studies of static and thermodynamic properties
under experimental conditions.}
In addition, we provide microscopic insights into 
the performance of the swap algorithm which should help
optimizing models and algorithms even further.
\end{abstract} 

\maketitle

\setlength{\onefig}{0.95\linewidth}
\setlength{\twofig}{0.45\linewidth}

\section{Introduction}

Computer simulations play an increasingly important role in elucidating
the nature of the glass transition because they allow particle-level resolution
of any relevant static or 
dynamic observable~\cite{BerthierBiroli2011}. While a similar spatial
resolution can now be achieved in experiments performed with colloids~\cite{Weeks2012},
less direct microscopic information is available from experimental studies 
of molecular liquids~\cite{Ediger2000}. Regarding timescales, however, colloidal experiments and 
computer simulations cover at best the first 4-5 decades of 
the dynamic slowing down of systems approaching a glass transition~\cite{brambilla},
whereas 12-13 orders of magnitude of glassy slowdown
can be analyzed in molecular liquids~\cite{Rossler2005}.
Therefore, the exquisite level of detail gained from
simulations in the description of the onset of slow dynamics 
concerns a dynamical regime which is separated from
experiments on molecular glasses by about eight orders of magnitude. 
The dichotomy between accessible length-scales and timescales 
is a major challenge for glass transition 
studies~\cite{richert,debenedetti,BerthierBiroli2011}.

There are several promising experimental advances which could
improve either the dynamic range of colloidal 
experiments~\cite{PADDYREF} or the spatial resolution in molecular 
supercooled liquids~\cite{illinois}. In addition, new protocols 
to prepare molecular glasses corresponding to even larger 
relaxation times are being developed~\cite{Ediger2007}. 
On the simulation front, the situation appears challenging,
as the increase in the time window accessible 
to computer simulations has been rather slow, 
amounting to a gain of about 3 orders of magnitude 
over the last 30 years~\cite{barrat,KABM,brambilla}, 
and this is mostly due to improvements in 
computer hardware. A rough extrapolation of this 
trend would pessimistically
suggest that it could take another 100 years for simulations
to close the gap with experimentally relevant thermodynamic 
conditions. 
The recent advent of graphic processing units and accelerators in the high-performance computing arena suggests that progress could be made at a faster pace if novel technologies become available. Exploiting them in the context of molecular simulations~\cite{Anderson_Lorenz_Travesset_2008,Colberg_Höfling_2011,Bailey_Ingebrigtsen_Hansen_Veldhorst_Bøhling_Lemarchand_Olsen_Bacher_Larsen_Dyre_2015,Anderson_Jankowski_Grubb_Engel_Glotzer_2013,Meyer_2013} requires nonetheless a substantial investment in code development and low-level optimization.

The above summary suggests that it is desirable to
develop alternative strategies, 
which do not simply rely on the brute force increase of computing power. A possible path is to take advantage 
of the flexibility offered by simulations and implement algorithms 
that simulate equilibrium material properties more efficiently~\cite{newman1999monte}. Several
such strategies have already been explored. A first line of research 
concerns the development of collective particle displacements 
to improve sampling 
efficiency~\cite{dress_cluster_1995,santen,bernard_prl_2011}.
This approach follows the method employed 
to study phase transitions in spin systems~\cite{swendsen,Binder}.
For instance, the event-chain Monte Carlo algorithm has proved useful in the study
of two-dimensional melting~\cite{bernard_prl_2011}, 
but its gain in efficiency for the three-dimensional dense fluids considered here 
is at most a factor of 40~\cite{isobe}, which remains insufficient to close the gap with experiments.
A crucial aspect for the efficiency of this approach is the choice of the correct
type of collective move, which still requires some a priori knowledge
of the relaxation path used by the system~\cite{Vink_2014}. 
This, however, is precisely one of the informations that remains to be understood in 
fragile glass-forming materials.

\begin{figure*}[!tbp]
  \centering
  \includegraphics[width=0.95\textwidth]{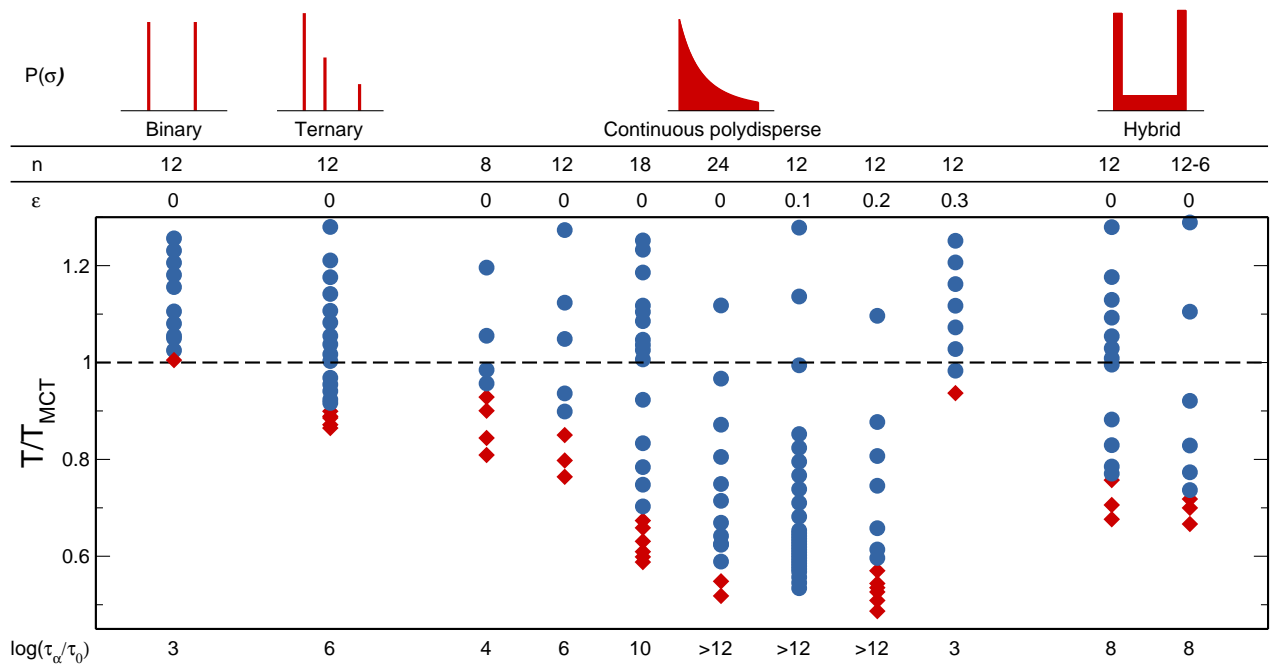}
  \caption{Summary of the results obtained for the eleven 
models studied in this work. The top row sketches 
the particle size distribution for each model, the next two lines specify
the pair potential and its additivity. Below, we use a 
temperature axis rescaled by the location of the mode-coupling 
crossover where blue 
points indicate equilibrium disordered fluid configurations,
red diamonds indicate instability towards crystalline or demixed states.
The bottom line indicates the estimated 
range of equilibrium relaxation times $\tau_\alpha$
that can be studied in stable
equilibrium conditions for each model, using $\tau_0$ as the 
relaxation time at the onset temperature~\cite{Sastry_Debenedetti_Stillinger_1998}.
We have constructed several models which remain stable and can be 
equilibrated deep in the temperature 
regime $T/T_{MCT} < 1$ that conventional simulation studies
are unable to penetrate, three of which allowing us to reach temperature
below the experimental glass transition, conventionally defined as 
$\tau_\alpha/\tau_0 = 10^{12}$.}
  \label{fig1}
\end{figure*}

A different simulation strategy is the replica-exchange 
technique, where simulations of the same system 
are conducted in parallel over a range of state points, and 
infrequent exchanges between neighboring state points 
are performed~\cite{Hukushima1996,Hukushima1998,yamamoto,odriozola}. 
The idea is that navigating through 
different state points would facilitate the crossing of 
large barriers in a complex free energy landscape, 
and indeed the technique was first developed to study 
spin glasses~\cite{Hukushima1996}. 
In dense fluids the reported speedup is again of about 
two orders of magnitude~\cite{yamamoto,odriozola}, 
with the additional drawback that the replica-exchange
technique scales very poorly with the number of particles 
and looses most of its 
efficiency for system sizes of thousands of particles, which are typically 
used in studies of the bulk glass transition~\cite{odriozola}. Therefore,
replica exchange works best for studies of equilibrium 
phase transitions in small systems, as confirmed in a 
series of recent 
studies~\cite{kob_probing_2013,ozawa_equilibrium_2015,Berthier13,berthier_jack_2015}.
Different algorithms such as Wang-Landau sampling~\cite{depablo-wang}
and population dynamics~\cite{machta} have also been employed in the 
context of glass studies.

The swap Monte Carlo algorithm is another 
longstanding simulation technique that has been used 
in computer studies of the glass transition. The algorithm 
was first introduced to study the equation of state of a 
non-additive hard sphere system~\cite{pastore} and 
later rediscovered in the context of the glass transition 
of a binary mixture of soft spheres~\cite{Grigera_Parisi_2001}.
The swap algorithm has since been mostly used in the glass 
context, for both binary mixtures~\cite{biroli_jcp_2008,cammarota_numerical_2009,cammarota_phase-separation_2010,cavagna_dynamic_2012}
and for continuously polydisperse systems~\cite{pronk2004melting,Fernandez_Mayor_Verrocchio_2007,bea}. For the binary
mixture of Ref.~\cite{bernu_soft-sphere_1987},  
the reported speedup in terms of equilibration times is a factor of 180,
independent of temperature~\cite{fernandez2006}. 
The glass-forming ability of this model is, however, poor
due to the appearance of ordered 
phases~\cite{brumer,Grigera_Parisi_2001,fernandezPhilo}. 
Little quantitative information is available concerning the efficiency of the 
swap algorithm for continuously polydisperse soft~\cite{brumer,Fernandez_Mayor_Verrocchio_2007,bea} and hard spheres~\cite{santenArxiv}.
In an effort to improve the stability of discrete mixtures, 
Gutierrez {\it et al.} recently introduced a ternary mixture 
of soft spheres to study the increase of a static 
length-scale~\cite{gutierrez}.
Very low temperatures were studied and a claim of a 10-decade
efficiency gain was made. We demonstrate below
that changing from a binary
to a ternary mixture indeed improves the
thermodynamic stability, but the claims made in 
\cite{gutierrez} do not 
resist our detailed analysis of the structure
and thermalization dynamics of the model. 
We will demonstrate that the efficiency gain for this model is much more 
modest and the 
accessible dynamical window is increased by about 2-3 orders of magnitude.

The aim of our work is to bring the swap algorithm to a
whole new level of performance. We present a systematic study 
of glass-forming ability and thermalization efficiency 
over a broad range of glass-forming models, 
varying the particle size distribution and the nature
of the pair interactions, while optimizing the swap Monte Carlo 
algorithm. Our main result, summarized in Fig.~\ref{fig1},
is the discovery that particular 
combinations of parameters yield both excellent glass-forming 
ability and a dramatic decrease of the 
computer time needed to obtain thermalized configurations at low
temperatures. This insight has already led to some new results on related phenomena, such as jamming~\cite{beyond} and the Gardner transition~\cite{gardner}.

As shown in Fig.~\ref{fig1},
we systematically change the size distribution,
using a variety of discrete and continuous mixtures, we vary the
softness of the pair repulsion, its additivity, and we add 
attractive forces. For each case, we determine both the temperature 
regime where the model is structurally unstable
(shown with red symbols) and the 
temperature regime where the disordered fluid states is stable at
equilibrium (shown with blue symbols). 
The vertical axis represents the temperature $T$, scaled by the 
location of the corresponding mode-coupling crossover, $T_{MCT}$.
Although somewhat arbitrary, this rescaling demonstrates the efficiency
of the thermalization because conventional computer simulations 
typically fail to reach equilibrium in the regime $T/T_{MCT} < 1$.
Despite the differences between systems, several of them can be 
thermalized in the supercooled liquid state at 
significantly lower temperatures than ordinary simulations. 
We demonstrate that this temperature regime corresponds, 
for some of these models, to a range of relaxation 
times of more than twelve decades, which implies that 
we can access in equilibrium a temperature regime that is even 
lower than the experimental glass transition temperature, $T_g$.
We show that this corresponds to a speed-up of the thermalization of about 
ten orders of magnitude at $T_g$.

The two key factors enabling such progress are 
the use of an appropriate size polydispersity to prevent 
both crystallization (when polydispersity is too small)
and phase separation (when it gets too large), and a particle
size distribution that allows for a large acceptance rate for
particle swaps, in turn leading to a fast thermalization 
and equilibrium sampling of phase space.

The outline of the article is as follows. Sec.~\ref{sec:sim} is dedicated 
to the simulation strategy and technicalities. Results for two families of 
systems (mixtures and continuous polydisperse systems) are reported and 
discussed respectively in Secs.~\ref{sec:mix} and~\ref{sec:poly}. We give 
a physical insight on swap dynamical relaxation and heterogeneities in 
Sec.~\ref{sec:relax}. Sec.~\ref{sec:design} deals with the introduction 
of a model designed to maximize the algorithm efficiency. Finally, 
Sec.~\ref{sec:discuss} presents our conclusions and offers 
further perspectives for future work.

\section{Details of the simulations}
\label{sec:sim}

\subsection{Algorithm, interactions and size distributions}
\label{sec:set}

We simulate systems of $N$ particles in a cubic box of side $L$ with periodic boundary conditions~\cite{allen}.
Throughout the paper we will compare results obtained from two kinds of 
simulation methods: 
standard Monte Carlo simulations in the canonical ensemble~\cite{FS01} 
and swap Monte Carlo simulations~\cite{pastore, Grigera_Parisi_2001}.
Both simulation algorithms involve the same 
displacement moves, in which we pick up one particle at random 
and attempt to translate it by a displacement vector 
randomly drawn in a cube of linear size $\delta l$. 
The move is accepted using the Metropolis acceptance rule, which ensures that 
detailed balance is obeyed at each temperature $T$. For each model, 
the typical jump length $\delta l$ is fixed to a fraction of the 
average particle diameter, which results in an acceptance 
rate ranging typically from about 60$\%$ at high temperatures 
to $30\%$ at low temperatures. This approach to simulating 
glass-formers has been validated by direct comparison with 
molecular dynamics results for the specific case of a binary 
mixture~\cite{Berthier2007}. 

In addition to displacement moves, during a swap Monte Carlo simulation we also attempt to exchange the diameters of two randomly chosen particles. The 
diameter exchange is again accepted based on the Metropolis criterion. 
At every Monte Carlo step, such a ``swap move'' is attempted with probability $p$. We emphasize that swap moves preserve detailed balance and thus guarantee an equilibrium sampling of phase space~\cite{newman1999monte}. In other words, despite the
``nonphysical'' nature of the swap moves (in an experiment, 
particles would not exchange their diameters spontaneously) 
the swap Monte Carlo dynamics 
enables a proper sampling of the equilibrium thermodynamic
properties of the model.
In previous implementations of the swap Monte Carlo, particle 
swaps were described as particles exchanging their positions, instead 
of their diameters~\cite{Grigera_Parisi_2001}. Both descriptions are of course fully equivalent, 
but our choice offers the advantage that single particle dynamics 
can be followed in time, because particles do not make arbitrarily large 
jumps during the swap moves.
Standard time correlation functions 
based on particle displacements can thus be measured
in swap and ordinary Monte Carlo simulations in the exact same way.
Dynamic measurements are a crucial tool to assess the thermalization
of our swap simulations, just as they are for standard simulations
of supercooled liquids. 
One Monte Carlo sweep is then defined as $N$ consecutive attempts 
to either displace or swap particles diameters, and one such sweep
will represent in the following our time unit.

In this work we study three different classes of systems, 
with particle size distributions as sketched in Fig.~\ref{fig1}. 
They are either discrete or continuous mixtures.
Discrete mixtures are characterized by a particle size distribution 
$P(\sigma)$ of the form
\begin{equation}
\label{eq:pmix}
P(\sigma) = \sum_{\alpha=1}^m x_\alpha \delta(\sigma-\sigma_\alpha), 
\end{equation}
where $m$ is the total number of components, $x_\alpha$ indicates 
the fractional composition of each species, and $\sigma_\alpha$ is the 
diameter of species $\alpha$.  
Within the class of continuously polydisperse systems, we focus on a 
specific kind of size dispersity, which scales as the inverse of the 
occupied volume:
\begin{equation}
\label{eq:pcon}
P(\sigma)= \frac{A}{\sigma^3}, \quad\quad \sigma\in[\sigma_{min}, \sigma_{max}],
\end{equation}
where $A$ is a normalizing constant and $\sigma_{min}$ and 
$\sigma_{max}$ are the minimum and the maximum diameter values, respectively. 
This functional form ensures that the volume fraction occupied by particles 
within a given bin size is constant. Such a scaling property has been shown to enhance glass-forming ability in discrete mixtures~\cite{ohern2015}, but we have not 
tested this hypothesis in great detail for the present systems.

Finally, we introduce a second type of continuous 
particle size distributions, which combine the salient features 
of both discrete and 
continuous mixtures. For this reason we call them ``hybrid'' distributions, 
see Fig.~\ref{fig1}. Mathematically, the distributions read  
\begin{equation}
\label{eq:phyb}
P(\sigma) = \sum_{\alpha=1}^m x_\alpha \theta(b_\alpha - |\sigma-\sigma_\alpha|),
\end{equation}
where $\theta(x)$ is the Heaviside function and $x_\alpha$ is 
defined as before. In this approach each component of the ``mixture'' 
is characterized by a flat particle size distribution of width $b_\alpha$. 
The goal is to construct models that combine advantages of both 
discrete mixtures, which are typically good glass-formers,
and continuous distributions, for which swap dynamics is very efficient. 

We quantify the degree of polydispersity 
of a system by the normalized root mean square deviation
\begin{equation}
\delta = \frac{\sqrt{\langle\sigma^2\rangle-
\langle\sigma\rangle^2}}{\langle\sigma\rangle},
\end{equation}
where the brackets indicate an average of the particle size 
distribution.
In the following, we will use $\langle\sigma\rangle = \int P(\sigma) 
\sigma d\sigma$ as the unit length for each studied model.

We model the interactions between two particles $i$ and $j$ 
via a soft repulsive pair potential of the type
\begin{equation}
\label{eq:pot}
v(r_{ij}) = \left(\frac{\sigma_{ij}}{r_{ij}}\right)^{n} + F(r_{ij}),
\end{equation}
where $n$ is an exponent controlling the softness of the 
repulsive potential, 
and $F(r_{ij})$ is a function that 
smooths the potential at the cutoff distance $r_{cut}$, beyond
which the potential is set to zero.
Unless otherwise specified, we use~\cite{gutierrez}
\begin{equation}
\label{eq:cut}
F(r_{ij}) = c_0  + c_2 \left(\frac{r_{ij}}{\sigma_{ij}}\right)^{2} + c_4 \left(\frac{r_{ij}}{\sigma_{ij}}\right)^{4} .
\end{equation}
The coefficients $c_0$, $c_2$, and $c_4$ 
ensure the continuity of the potential up to the 
second derivative at the cutoff distance $r_{cut}=1.25\sigma_{ij}$.
Additionally, we also studied a polydisperse model where particles interact with the Lennard-Jones potential
\begin{equation}
\label{eq:LJ}
v(r_{ij})= \left(\frac{\sigma_{ij}}{r_{ij}}\right)^{12} - 
\left(\frac{\sigma_{ij}}{r_{ij}}\right)^{6} +  c_{LJ},
\end{equation}
for which we simply cutoff and shift the pair potential 
at the cutoff distance $r_{cut} = 2.5 \sigma_{ij}$.

Finally, to ensure a high structural stability in our models, 
we introduce a generalized non-additive interaction rule for 
the cross diameters $\sigma_{ij}$ in the pair interaction, which reads
\begin{equation}
 \label{eq:sigNA}
\sigma_{ij}= \frac{\sigma_i+\sigma_j}{2}(1-\epsilon |\sigma_i-\sigma_j| ).
\end{equation}
Systems characterized by $\epsilon=0$ 
and $\epsilon\neq0$ will be referred to as additive and non-additive systems, 
respectively. Non-additivity is another ingredient which has been
widely used to enhance glass-forming ability in simple binary models~\cite{Kob1994} and is a consequence of the band structure of the electronic density of states in metallic alloys~\cite{Hausleitner_Hafner_1992}. Physically, the non-additive rule in Eq.~\eqref{eq:sigNA} implies that particles with identical 
diameters interact as before, but that small and large particles 
can have a larger overlap than for additive systems.
  
\subsection{Physical observables}
\label{sec:phys}

In this section we introduce the basic observables used to characterize the structure and dynamics of the studied models. We will use them to monitor the equilibration and the stability of the fluids under supercooled conditions and to quantify and compare 
the degree of thermalization achieved by both standard and swap simulations.

We systematically compute the structure factor~\cite{HM90}, 
\begin{equation}
S(k)=\frac{1}{N}\langle \rho_{\bf k} \rho_{-{\bf k}}\rangle,
\end{equation} 
where $\rho_{\bf k}$ is the Fourier transform of 
the microscopic density at wavevector ${\bf k}$. 
The behavior of $S(k)$ at small wave-number provides information on 
possible long-range density fluctuations and will be 
checked to identify signals of instability of the homogeneous fluid.
Since we deal with size-disperse systems, we compute partial structure factors associated to each subpopulation. 
In the case of continuously polydisperse systems, we group particles of comparable size into families labeled by an index $\alpha$, for which we compute the 
partial structure factor $S_{\alpha\alpha}(k)$. A strong increase 
of $S_{\alpha\alpha}(k)$ at small $k$ values is associated to 
phase separation or demixing, and we have monitored 
this quantity systematically in our models.

Beside particle demixing, the main instability to be overcome 
is of course crystallization. 
To detect the presence of crystalline local order, we measure the 6-fold 
bond-orientational order parameter~\cite{tanaka}
\begin{equation}
\label{eq:boo}
Q_6=\left\langle\frac{1}{N}\sum_i\sqrt{ \frac{4\pi}{13}\sum_{m=-6}^{6}\left|\frac{1}{N_b(i)}\sum_{j=1}^{N_b(i)}Y_{6m}(r_{ij})\right|^2 }\right\rangle,
\end{equation}
where $Y_{6m}(r_{ij})$ are spherical harmonics. The sum over $1<j<N_b(i)$ runs 
over the neighbors of particle $i$ in a sphere of radius corresponding to the 
minimum of the distribution function 
of rescaled inter-particle distances,  $g(r_{ij}/\sigma_{ij})$. We inspect 
the time and temperature variation of $Q_6$, as well as that of the potential 
energy $e$, to check whether the systems are stable against crystallization.

We provide a systematic characterization of both self and collective dynamics of the models. This enables us to quantify the degree to which swap simulations enhance thermalization compared to standard Monte Carlo. We note that while standard Monte Carlo dynamics~\cite{Berthier2007} can be used to mimic overdamped Brownian dynamics, as appropriate for a colloidal suspension~\cite{brambilla}, the microscopic dynamics of swap simulations is not physical. We emphasize however that particles' trajectories remain well-defined, because the swap moves only exchange the particle diameters, not their positions. Thus, even though the microscopic dynamics is nonphysical, time-dependent correlation functions still quantify the timescale over which individual particles 
diffuse (for self-correlation functions) and over which the 
density fluctuations relax (for collective correlations). Time correlation
functions will be used in the following to determine whether the system has been 
efficiently thermalized at a given state point.

We characterize the single particle dynamics 
through the self-part of the intermediate scattering function
\begin{equation}
\label{eq:self}
F_s(k,t)= \langle  f_s(k,t) \rangle =  \left\langle 
\frac{1}{N}\sum_j e^{i\textbf{k}\cdot[\textbf{r}_j(t)-\textbf{r}_j(0)]} \right\rangle ,
\end{equation}
where the wavenumber $k$ corresponds to the first peak of the total 
structure factor $S(k)$.
Notice that since the particle diameter changes during the course of the
simulations, the sum in Eq.~(\ref{eq:self}) runs over all particles, 
the distinction between large and small particles being 
immaterial. 
The structural relaxation time $\tau_\alpha$ is then defined as the value 
at which $F_s(k,\tau_\alpha)=e^{-1}$, following common practice. We use the 
relaxation time $\tau_\alpha$ measured for standard Monte Carlo 
simulations to locate the mode-coupling crossover at $T=T_{MCT}$, 
which we take as a relevant temperature scale for computer simulations.
{In order to obtain $T_{MCT}$ we fit the standard dynamics (without swap) 
in the interval $\tau_0<\tau_\alpha<10^3\tau_0$}
with a power law divergence~\cite{goetze}, 
\begin{equation}
\label{eq:MCT}
\tau_\alpha\propto (T-T_{MCT})^{-\gamma}.
\end{equation}
When discussing the dynamics of our models, we will also use 
other functional forms to describe the temperature evolution of the 
relaxation time. A well-known functional form is 
the Vogel-Fulcher-Tamman (VFT) law~\cite{BerthierBiroli2011}, 
\begin{equation}
\label{eq:VFT}
\tau_\alpha = \tau_\infty \exp \left( \frac{A}{T-T_0}  \right),
\end{equation}
where $\tau_\infty$, $A$ and $T_0$ are fitting parameters. 
Because this functional form describes a dynamic singularity 
at a finite temperature $T=T_0$, 
it produces a very steep temperature dependence. 
A less pronounced temperature dependence is obtained
with the parabolic law~\cite{Elmatad2009}, 
\begin{equation}
\label{eq:PARAB}
\tau_\alpha = \tau'_\infty \exp \left[ A' \left(\frac{1}{T}-\frac{1}{T_1}\right)^2 \right],
\end{equation} 
where $\tau_\infty'$, $A'$ and $T_1$ are again free parameters. 
Notice that no dynamic singularity is predicted from 
Eq.~(\ref{eq:PARAB}), since $T_1$ captures the onset of slow
dynamics and not the divergence of the relaxation time at low 
temperature. A final form that we use is the Arrhenius law, 
\begin{equation}
\label{eq:ARRH}
\tau_\alpha = \tau''_\infty \exp \left( \frac{A''}{T} \right),
\end{equation}  
with $\tau''_\infty$ and $A''$ two fitting parameters. 

Using these functional forms will be useful below to 
estimate the range of relaxation times that swap dynamics allows
us to access. Our analysis shows that the VFT 
law presumably overestimates the growth of the relaxation time whereas
the Arrhenius law underestimates it, the parabolic law 
falling somewhat in-between. Thus the combination of 
all three fitting functions provides an estimate of 
the actual physical behavior and a sensible confidence interval in low temperature extrapolations. 

The relaxation of collective density fluctuations is measured via the time-dependent overlap function
\begin{equation}
\label{eq:coll}
F_o(t)=\left<\frac{1}{N}\sum_{i,j}\theta(a-|\textbf{r}_i(t)-\textbf{r}_j(0)|)\right>,
\end{equation}
using a cutoff distance $a=0.3$. This quantity provides similar information as the coherent intermediate scattering function at wavevector $k = 2 \pi /a$,
but is computationally more advantageous because 
it presents much smaller statistical fluctuations.
From this function, we define a relaxation time $\tau_o$ for the 
decorrelation of collective density fluctuations, such that 
$F_o(\tau_o)=e^{-1}$. 

For selected models we computed a number of additional static and 
dynamical observables with the aim of understanding 
microscopic processes taking place during the swap Monte Carlo 
simulations. These 
more specific observables are described later in Sec.~\ref{sec:relax}.

\subsection{Efficiency of swap moves}

Because the swap Monte Carlo is conceptually very simple, there 
are very few parameters that can be adjusted to optimize its 
efficiency. We discuss how to achieve maximal efficiency 
in the present section. 

The extent to which swap moves accelerate the sampling of 
configuration space during a Monte Carlo 
simulation must depend on the frequency used to 
attempt such moves, which is given by the probability $p$.
There are two obvious limiting cases. For $p=0$ one recovers the dynamics of a standard Monte Carlo simulation without swap moves.
For $p=1$, instead, only swap moves are attempted and 
the particle positions are never updated, so that by construction
structural relaxation cannot take place.
The optimal choice for $p$ is thus the one that minimizes the 
structural relaxation time $\tau_\alpha$ of the system with respect to $p$.

\begin{figure}
\centerline{\includegraphics[width=\onefig]{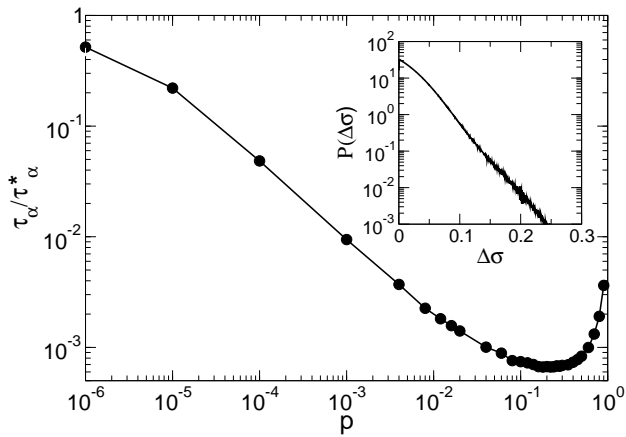}}
\caption{The relaxation time $\tau_\alpha$ as function of the swap 
attempt probability $p$, normalized by $\tau^*_\alpha$, its value 
for standard Monte Carlo simulations when $p=0$. A broad minimum 
indicates that $p \approx 0.20$ optimizes the efficiency 
of the swap Monte Carlo algorithm.  
The inset shows the probability distribution of swap acceptance 
as a function of the diameter difference $\Delta \sigma =
|\sigma_1 - \sigma_2|$ between the particles for which 
the swap move is attempted. 
The system is a soft repulsive system of non-additive particles 
studied in Sec.~\ref{sec:nadd}, with $\epsilon=0.2$ and $T=0.101$.}
\label{fig3}
\end{figure}

We illustrate the optimisation procedure for a 
continuously polydisperse particles system interacting via a soft 
repulsive potential as in Eq.~(\ref{eq:pot}) 
with $n=12$, with a non-additivity $\epsilon=0.2$. This model 
is further discussed in Sec.~\ref{sec:nadd}. 
The general trend found for this model is representative of the three classes of systems we investigated and is shown in Fig.~\ref{fig3}, where we report the structural relaxation time of the system versus $p$ at a constant 
temperature, $T=0.101$. We normalized the relaxation time by the corresponding value in the absence of swap moves at $p=0$. At this particular temperature, we observe that the structural relaxation time becomes almost three orders of magnitude faster compared to standard dynamics already for very small values of $p$, i.e. $p$ of the order of a few percents. We observe a relatively broad minimum around $p \approx 0.2$ before $\tau_\alpha$ starts to grow again and diverges for $p=1$ as $\tau_\alpha \sim (1-p)^{-1}$,
when particles stop diffusing for the trivial reason mentioned above. 
From such a graph, we deduce that $p \approx 0.2$ is the optimal 
value for the probability to perform swap moves. We find that 
this value is fairly robust when temperature is changed or across different 
models, which presumably stems from the fact that the minimum reported 
in Fig.~\ref{fig3} is relatively flat. Another remarkable feature of this 
figure is the very steep decrease of $\tau_\alpha$ observed for even very small
values of $p$ suggesting that even a fairly small amount of swap 
moves is in fact sufficient to facilitate enormously the structural 
relaxation of the system. 

Efficiency considerations should also take into account the CPU time needed to perform a swap move as opposed to a standard displacement. 
An attempt to swap diameters entails the computation of the local energy variation between the new and the old configuration for two particles, which is twice what is needed for an ordinary 
displacement move involving only one particle. 
However, we found that the optimum value for $p$ barely changes even when taking this additional effect into account. In terms of CPU time, one MC sweep with $p=0.2$ takes only $20\%$ longer on average than a standard sweep with 
$p=0$. This should be contrasted with the orders of magnitude of
gain achieved in terms of structural relaxation time.

Another major advantage of the swap Monte Carlo algorithm is that 
both its implementation and its efficiency are insensitive to the 
number of particles in the system, $N$. This contrasts strongly with 
the replica exchange method, which scales very poorly with $N$~\cite{Hukushima1996,Hukushima1998,yamamoto,odriozola}. 

In general, the acceptance ratio of Monte Carlo moves decreases upon lowering temperature or increasing the density. Similarly, the acceptance ratio of swap moves decreases when the size difference $\Delta \sigma=|\sigma_1-\sigma_2|$
of the two selected particles increases, because a large particle will not 
easily fit into the hole occupied by a small one. As will be clear in the following, the efficiency of swap moves is highest in continuously polydisperse systems, where the diameter difference between any two particles can be arbitrarily small. In these systems, it is pertinent to avoid attempting exchanges when 
$\Delta \sigma$ is too large because the swap move is then essentially 
always rejected~\cite{brumer}. This point is illustrated in the inset of Fig.~\ref{fig3}, 
where we show $P(\Delta \sigma)$, the probability distribution of 
acceptance rates for swap moves between pairs of particles 
with a size difference $\Delta \sigma$ for the same parameters
as in the main frame of the figure. 
We notice that the acceptance rate decays exponentially fast with $\Delta \sigma$ and becomes vanishingly small when $\Delta \sigma \gtrsim 0.25$.  
Following Ref.~\cite{brumer}, we therefore disregard swaps between particles with a diameter difference larger than a certain cutoff. We choose here $\Delta \sigma_{max}=0.2$, which we found to be a reasonable trade-off. 
We implement this threshold value in a way that preserves 
detailed balance. In practice, we
always choose two particles at random, but we directly reject the swap 
without evaluating any energy difference if $\Delta \sigma$ exceeds 
the chosen cutoff value. 

\subsection{Equilibration and metastability}
\label{sec:equi}

Simulations of glass-forming liquids must be long enough to ensure equilibrium sampling of the observables of interest and yet short enough to avoid crystallization or more complex forms of structural ordering. Simple models such as binary mixtures or weakly polydisperse systems have been shown to crystallize over sufficiently long times~\cite{brumer,bea, Fernandez_Mayor_Verrocchio_2007,wilding2010,sollich2010,Ingebrigtsen_Tanaka_2015}. Computer simulations of glass-forming materials 
thus always represent a narrow compromise between those two limits
that both need to be addressed carefully.

These issues become particularly severe when employing enhanced sampling algorithms, such as swap Monte Carlo moves which are precisely constructed to 
promote a more efficient exploration of phase space. 
For instance, crystallization of two-component mixtures of repulsive spheres 
has been reported in swap Monte Carlo 
simulations~\cite{brumer, fernandezPhilo, cavagna}.
The ground-state of polydisperse repulsive particles was studied both with swap Monte Carlo~\cite{Fernandez_Mayor_Verrocchio_2007} and in the semi-grand canonical ensemble~\cite{wilding2010,sollich2010}. These studies found that for sufficiently high polydispersity the stable structure is a fractionated crystal, where the system presents multiple crystals each involving a fraction of the system overall particle size distribution.
However, as noticed in \cite{wilding2010}, the free energy cost of forming an interface between distinct phases is generally high and crystallization may be 
difficult to observe in practice.

The general conclusion to be drawn from these earlier works is that 
a model considered as a good glass-forming system when studied using 
conventional simulations techniques may turn out to be a very poor model
when using an enhanced simulation technique that is able to probe 
a much wider 
range of temperatures. Indeed, we find 
that many previously studied types of glass-forming models 
do not withstand basic stability criteria when the swap technique is applied, forcing us to develop novel numerical models in addition to the 
optimization of the swap Monte Carlo method.  

To make a consistent comparison of the glass-forming ability of the studied models in standard and swap simulations, we follow a 
rigorous and identical equilibration protocol for all our models.
First, we obtain static and dynamical properties of the system by means of standard Metropolis Monte Carlo simulations in the $NVT$ ensemble~\cite{FS01}.
From these simulations we extract the average potential energy 
value, the structure factors and the structural relaxation times. For each model we determine 
$T_{MCT}$ using Eq.~\eqref{eq:MCT}, which will serve as a reference 
temperature scale to compare the degree of supercooling across 
different models. This is not an ideal choice, but it offers the advantage 
that extrapolation of the relaxation times to low temperatures 
is not needed.

Swap Monte Carlo simulations start from a configuration equilibrated at the onset temperature $T_{o}$~\cite{Sastry_Debenedetti_Stillinger_1998}, followed by an instantaneous quench to the target temperature $T$. The following criteria are used to determine whether the 
system has reached equilibrium at $T$.
First, we monitor the potential energy $e$ per particle. We inspect both its instantaneous value as a function of time, $e(t,T)$, to detect aging, as well as its time average as a function of temperature 
$\langle e \rangle(T)$, to detect possible discontinuities or change of slope
in the equation of state of the liquid. 
Second, we ensure that the total mean-squared displacement $\Delta r^2(t) = \langle 1/N \sum_i [\textbf{r}_i(t)-\textbf{r}_i(0)]^2\rangle$  has reached a value at least larger than $6$. This specific value is relatively immaterial 
as this criterion only conveniently guides us between state points where 
particle displacements
are large over the numerical time window, from those where particle
dynamics is essentially arrested. 
Finally, we look at the self-incoherent scattering function. For this quantity, we check, within statistical fluctuations, both the absence of aging and the complete decorrelation to zero at long times.
Once equilibration has been reached, we perform a first set of simulations to 
obtain a rough estimate of the structural relaxation time $\tau_\alpha$ 
in the presence of swap moves. After this is done, the system is 
simulated over a total of $200\tau_{\alpha}$ to measure 
static and dynamic properties over a sufficiently wide time window. 
Notice that this thermalization procedure is rather demanding and 
thermalization thus requires that we are able to perform simulations 
over a time window which is 2 orders of magnitude longer than the 
structural relaxation time. We emphasize that such procedure is 
not specific to the presence of the swap moves. We think that 
simulation of supercooled liquid should follow similar strict
rules to claim that equilibration as well as a proper sampling 
of phase space has been achieved.

Almost every system we simulated eventually displays some form of
structural instability at sufficiently low temperature, such as
nucleation of an ordered phase or long wavelength density fluctuations
and demixing.  These instabilities are detected via the observables
introduced in the previous section. To decide whether a given state
point remains in a metastable disordered fluid state, we use the
following criterion.  We perform five independent simulations along
the lines described above. If at least one among the five samples
shows an instability over a time window of $200 \tau_\alpha$, then we
classify this state point as unstable and that the system is a poor
glass-former at this temperature. The precise meaning of
``instability'' is system-dependent and will be specified in
each studied case in
Secs.~\ref{sec:mix} and~\ref{sec:poly}. This criterion is 
rather strict, as there could still be room for achieving
thermalization and performing equilibrium sampling while avoiding
structural instability, but this is dangerous as fluctuations related
to ordering could interfere with the physics of the metastable
disordered state. In addition, we find that when a state point is
deemed unstable using our criterion, then lower temperatures are also
unstable and the instability rapidly becomes so severe that further
studies can not be safely performed.  Thus, changing the details of
our criterion would simply shift the range of ``metastable''
temperatures by a very small amount and our conclusions would not be
affected.

\begin{figure}
\centerline{\includegraphics[width=\onefig]{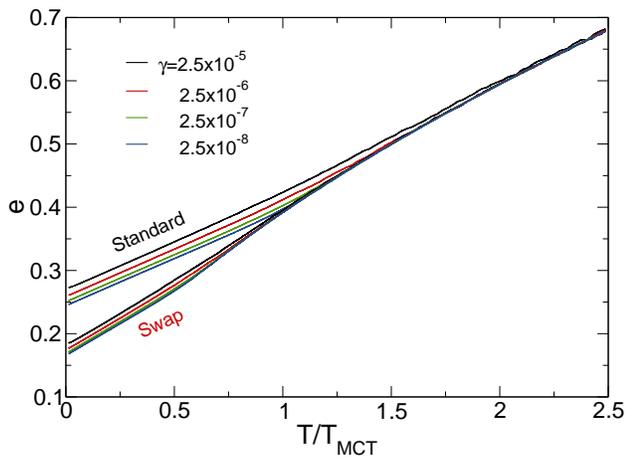}}
\caption{Potential energy per particle
in simulations with different cooling rates $\gamma$ using 
both standard and swap Monte Carlo. 
Both dynamics produce consistent results at high temperatures, but the
swap dynamics remains at 
equilibrium down to much lower temperatures than the standard one. 
The system is the soft repulsive system of non-additive particles 
studied in Sec.~\ref{sec:nadd}, with $\epsilon=0.2$.}
\label{fig2half}
\end{figure}

We directly compare the equilibration process between standard and 
swap simulations in Fig.~\ref{fig2half}.
Starting from a high temperature configuration we progressively 
quench the system 
down to zero temperature using a constant
cooling rate $\gamma=dT/dt$ with values changing logarithmically 
over a broad interval
 ($2.5\times10^{-5},2.5\times10^{-9})$.
We further average our results using ten independent initial configurations.
For standard Monte Carlo simulations, we retrieve 
the expected behaviour where departure from the equilibrium equation state 
arises at lower temperature for lower cooling rate, as
signalled by a rate dependence of the energy. 
Using swap simulations, we obtain the very same equation of state 
at equilibrium, and a similar rate-dependent behaviour at low temperatures. 
The major difference between the two sets of simulations is that 
swap simulations clearly fall out of equilibrium at considerably lower
temperatures than ordinary Monte Carlo simulations. The agreement 
of the two sets of curves when they both probe equilibrium is an indication
that swap dynamics has been correctly implemented and and provides the 
correct sampling of phase space. The second information gained from 
this set of data is the clear indication that the swap Monte Carlo 
algorithm extends the regime where equilibrium studies are possible
by a large amount and is able to 
produce highly stable glass configurations. 

\begin{figure}
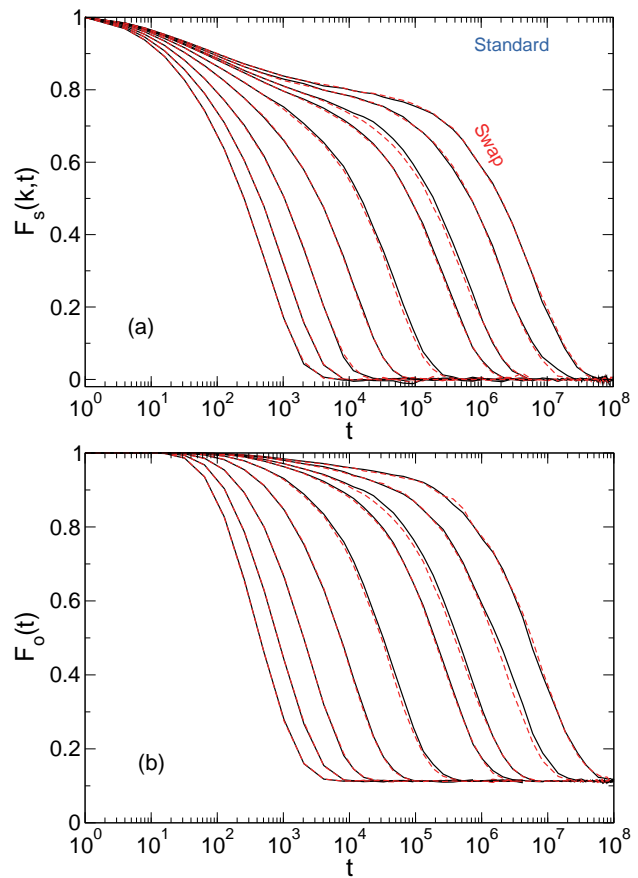

  \centering
  \includegraphics[width=\onefig,clip]{fig3a.eps}
  \includegraphics[width=\onefig,clip]{fig3b.eps}
  \caption{\label{fig2} 
Swap dynamics in a soft repulsive system of non-additive particles 
studied in Sec.~\ref{sec:nadd} with $\epsilon=0.2$.
(a) Self-incoherent scattering $F_s(k,t)$ computed respectively on the 
first (full lines) and second (dashed lines) half 
of the simulation run. Results for standard dynamics without swap for the 
lowest temperature are shown with a dotted line.
(b) Collective overlap correlation function $F_o(t)$. 
In both panels, temperatures are $T=0.25$, $0.175$, $0.125$, $0.092$, 
$0.075$, $0.065$, $0.062$, $0.058$, $0.0555$, from 
left to right.
Swap Monte Carlo simulations fully decorrelate 
both single particle and collective 
density fluctuations in a regime where standard Monte Carlo simulations
may be fully arrested.}
\end{figure}  

To illustrate our equilibration protocol, we show in Fig.~\ref{fig2}(a) 
the incoherent scattering function for the same system 
as in Fig.~\ref{fig3} (see also Sec.~\ref{sec:nadd}) 
evaluated over the first and the second halves of the simulation at various temperatures. As we can see, the two sets of curves agree within statistical uncertainty over a wide range of temperatures, demonstrating the absence of aging. For the lowest temperature at which thermalization with swap moves is achieved, we show the corresponding relaxation function obtained 
without swap, which quickly decays to a plateau that extends over the
last 6 decades of the simulation. This shows that without swap moves, 
the dynamics is fully arrested at these low temperatures, 
and no equilibrium simulations can presently be performed in this regime with conventional computational techniques.
 
In Fig.~\ref{fig2}(b), 
we show the collective overlap function, 
Eq.~\eqref{eq:coll}, which decorrelates to a density-dependent plateau 
at long times, as expected in ergodic equilibrium simulations. 
The two plots of Fig.~\ref{fig2} underline the fact that swap Monte Carlo
simulations fully decorrelate both single particle and collective 
density fluctuations in a regime where standard Monte Carlo simulations
may be fully arrested and therefore represent an efficient and reliable 
method to sample the configuration space.

\section{Results for discrete mixtures}
\label{sec:mix}

\subsection{Binary mixtures}
\label{sec:bin}

Simple binary mixtures of repulsive spheres were the first computer models 
for supercooled liquids simulated using the swap Monte Carlo method~\cite{pastore,Grigera_Parisi_2001}.
Here, we focus on the ``historical'' 50:50 binary mixture introduced long ago 
by Bernu {\it et al.}~\cite{bernu}, which has been frequently used since
its introduction.
The pair interaction is given by Eqs.~\eqref{eq:pot} and \eqref{eq:sigNA} with $\epsilon=0$ and $F(r_{ij})= c_{\alpha \beta}$, where $\alpha,\beta=A,B$ are species indexes. The size ratio is $\frac{\sigma_A}{\sigma_B}=1.2$, 
resulting in a polydispersity $\delta = 9.1 \%$. 
The potential is cutoff and shifted at a distance 
$r_{cut}=\sqrt{3}$, a specific value which was often used 
in previous studies~\cite{Grigera_Parisi_2001,brumer,cavagna}. 
We simulate $N=1024$ particles at the number density $\rho=1$. 

\begin{figure}
  \includegraphics[width=\onefig]{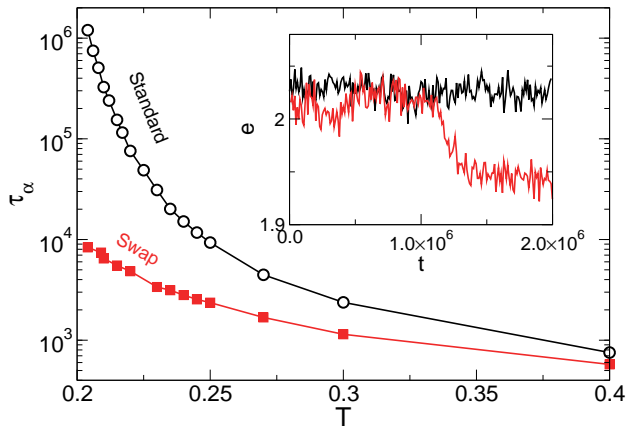}
  \caption{Relaxation times $\tau_\alpha$ of standard (black empty points) 
and swap (red full squares) simulations for the binary mixture
of soft repulsive spheres studied in Sec.~\ref{sec:bin}.
The speedup offered by the swap moves is obvious, but the system is unstable
below $T = 0.202$ where it crystallizes, and 
temperatures below the mode-coupling temperature $T_{MCT} = 0.199$ cannot
be studied. The inset 
shows a time series of the potential energy for standard (black) 
and swap (red) simulations at $T=0.2$, showing that 
crystallization is easily observed when swap moves are introduced.}
  \label{fig4}
\end{figure}

As already demonstrated before~\cite{cavagna}, swap moves help to accelerate sampling in this system, even though their acceptance rates is relatively low, of order $a\sim 10^{-2}$. We confirm this finding in
Fig.~\ref{fig4} where we compare the structural relaxation time
$\tau_\alpha$ measured during standard and swap Monte Carlo simulations. 
Over the range of temperatures at which the system can be equilibrated according to our criteria (see Sec.~\ref{sec:equi}), swap moves result in a speedup of about $2$ orders of magnitude at the lowest 
temperature. Notice that contrary to published analysis~\cite{fernandez2006}, 
we find that the efficiency of the swap over the standard Monte Carlo method is 
strongly temperature-dependent, and efficiency 
increases rapidly as temperature decreases.

Unfortunately, however, the temperature range 
that can be analyzed with this system 
does not change dramatically when swap moves are introduced. 
In fact, even using standard MC, the system crystallizes at the lowest studied temperature and becomes unstable when $T<0.202$, which is marginally 
larger than the location of the mode-coupling crossover, 
$T_{MCT} \approx 0.199$. Notice that earlier, incorrect determinations of the mode-coupling crossover temperature of this system have 
misleadingly suggested that temperatures well below 
$T_{MCT}$ could be simulated with this system. In reality, $T_{MCT}$ 
represents the lowest temperature that can be safely studied, 
swap moves merely providing a more efficient way of producing 
thermalized configuration in the temperature regime $T > T_{MCT}$. 
In other words, swap MC accelerates the dynamics of the system 
but does not allow the exploration of a temperature regime that 
is not already accessible with ordinary simulations.

In the inset of Fig.~\ref{fig4} we show the time series of the potential energy of a sample at $T=0.2$, where rapid crystallization is observed when swap 
dynamics is employed.  
We note that crystallization in this model is well-documented and has been 
studied in detail in small samples~\cite{brumer,Berthier2012}. 
Since complex strategies would be needed to detect and filter out 
crystallized configurations~\cite{cavagna} already near the 
mode-coupling crossover, we conclude that this ``historical'' model 
can indeed be efficiently 
simulated using swap Monte Carlo but is too poor a glass-former to 
fruitfully explore novel physical regimes. 

Within the realm of simple binary mixtures, it is difficult 
to make further progress using swap Monte Carlo 
because to increase the structural stability of the 
system one would need to increase the size ratio (for instance using the 
more stable $\sigma_A/\sigma_B = 1.4$ well-studied model), but this 
would imply that the already very low acceptance rate for swap moves would
become vanishingly small and swap would thus not be a useful method.
Therefore, the trade-off between stability and swap efficiency 
leaves very little room for a drastic improvement of simulation techniques when
binary mixture models of glass-formers are used.
Another option is to introduce non-additivity in the interactions, as 
for instance in the classic Kob-Andersen mixture~\cite{Kob1994}. 
Rather than for binary mixtures, we will explore this possibility for a 
different family of models based on continuously polydisperse particles 
(see Sec.~\ref{sec:poly}).

\subsection{A ternary mixture}
\label{sec:tern}

Given the limits demonstrated above for binary mixtures,
a natural strategy is to increase the number of components 
in the model. Adding more chemical components is indeed a commonly used
method to improve the glass-forming ability of metallic alloys. 
In addition, by increasing the number of components, one can
simultaneously increase the polydispersity, and thus the glass-forming 
ability of the model, while preserving the swap efficiency
by introducing particles with size ratios that are small enough for 
swap moves to be frequently accepted.

This strategy was recently followed in Ref.~\cite{gutierrez}, 
where a ternary mixture of soft spheres was introduced and studied
using swap Monte Carlo dynamics. 
The potential used in that work can be cast in the form of Eq.~\eqref{eq:pot}
with a softness parameter $n=12$ and 
$F(r_{ij})$ as in Eq.~\eqref{eq:cut}, with 
a cutoff distance $r_{cut}=1.25\sigma_{ij}$. 
We simulated systems with $N=1500$ particles at the number density 
$\rho=1.1$, as in the original version of the model~\cite{gutierrez}.
The size ratio between two species is 
$\frac{\sigma_{A}}{\sigma_{B}}=\frac{\sigma_{B}}{\sigma_{C}}=1.25$, 
which is slightly larger than for the binary mixture 
studied above in Sec.~\ref{sec:bin}, and compositions 
$x_A = 0.55$, $x_B = 0.30$, and $x_c= 0.15$, which 
ensures that all species roughly occupy the same fraction of the total
volume.
The corresponding size polydispersity is $\delta \approx 17 \%$, 
and so we can expect a smaller tendency for the system to crystallize.
Simultaneously, we also expect the 
acceptance of the swap moves to be much smaller than for the binary mixture.
In agreement with Ref.~\cite{gutierrez}, we find that the acceptance 
rate for swaps is of the order $a \sim 10^{-5}$ at low temperatures.
To speed up the simulations, we therefore only attempt swap 
moves between species $(A, B)$ and $(B, C)$ separately, because the probability 
of accepting swaps between pairs of $(A, C)$ particles is 
negligible.

Despite the low acceptance rate, it was claimed in 
Ref.~\cite{gutierrez} that swap moves allow for a dramatic 
speedup of the thermalization in this model. In the reduced 
units described above, we locate the mode-coupling crossover 
temperature near $T_{MCT} \approx 0.288$, and Gutierrez 
{\it et al.} claim to have achieved thermalization  down to 
$T=0.22 \approx 0.76 T_{MCT}$.
Based on dynamic scaling arguments, they estimate that the relaxation
time at $T=0.22$ is $\tau_\alpha / \tau_0 \approx 10^{15}$, 
where $\tau_0$ is the value of the relaxation time 
near the onset temperature $T_0$. Thus, the claim is that 
swap Monte Carlo provides an increase in the 
accessible window of relaxation times of about 10 orders of magnitude
as compared to standard molecular dynamics 
simulations~\cite{gutierrez}.

\begin{figure}
  \includegraphics[width=\onefig]{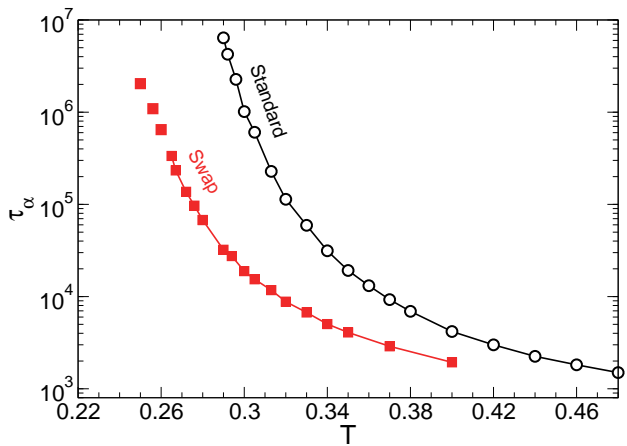}
  \caption{Relaxation times $\tau_\alpha$ of standard (black empty points) 
and swap (red full squares) simulations for the ternary mixture
of soft repulsive spheres studied in Sec.~\ref{sec:tern}.
The speedup offered by the swap moves is obvious, but the system is unstable
below $T = 0.26 \approx 0.9 T_{MCT}$ where it demixes and crystallizes. 
Disconnected squares are a rough
estimate of $\tau_\alpha$ obtained using short simulations 
in the unstable region. Stable and equilibrated
states can be accessed down to $T \approx 0.26 < T_{MCT} =0.288$,
extending the dynamic range by about 2 orders of magnitude
as compared to standard simulations.}
  \label{fig5}
\end{figure}

We have repeated and extended 
these simulations using standard and swap Monte Carlo
dynamics. In Fig.~\ref{fig5} we present the temperature evolution 
of the structural relaxation times 
for both these dynamics. We confirm that despite the 
very low acceptance rate of the swap moves, the speedup 
of the dynamics produced by these rare swaps is important.
For instance, at the lowest temperature simulated by standard 
Monte Carlo, $T=0.29$ 
the relaxation time is reduced by a factor of about $10^2$ when swap moves 
are introduced. Following the evolution of $\tau_\alpha$ using 
swap moves, we find however that $\tau_\alpha$ becomes 
too large to be accurately measured for $T \leq 0.24$ and 
particles in fact barely move over  the 
entire simulation performed at $T=0.22$. We conclude therefore that 
our swap Monte Carlo simulations fail to thermalize the model for $T \leq 0.24$.
In Ref.~\cite{gutierrez}, thermalization was tested by reweighting the probability distribution functions 
of the potential energy. We could reproduce this thermalization test in our work, thus demonstrating that this test fails to detect the lack 
of thermalization and proper sampling for the lowest studied 
temperatures. Measuring the structural relaxation time and the relaxation dynamics
is thus a more accurate and more discriminating thermalization test 
than techniques based on global static observables only.

In addition to the lack of thermalization at low temperatures, 
we also find signatures of structural instability of the fluid 
at even higher temperatures, $T \leq 0.26$. 
Below this temperature, our criteria  
for absence of crystallization or demixing are no longer fulfilled, 
and the system is eventually unstable within the window 
of $200 \tau_\alpha$ that we use to assess stability. Using shorter 
time windows before the system crystallizes, 
we obtain a rough estimate of the relaxation time $\tau_\alpha$
in the unstable regime, and show these results as disconnected squares in Fig.~\ref{fig5}. 
In Ref.~\cite{gutierrez}, the glass-forming ability of the model
was not discussed but there may be evidence of ordering in the reported
peak of the specific heat. An alternative reason for 
the absence of ordering in the data of Ref.~\cite{gutierrez} 
is that the performed simulations covered a smaller time window
of about $10^6-10^7$ Monte Carlo sweeps, whereas we simulate 
up to $10^9$ sweeps in our work. Of course, preventing ordering 
through shorter simulations implies that thermalization becomes more
difficult to achieve, and an accurate sampling of phase space 
is then problematic.

Comparing the stable results for the ternary mixture to the 
ones of the binary mixture in Fig.~\ref{fig4}, it is clear that 
the efficiency of swap Monte Carlo is essentially preserved, and that 
thermalization and metastability of the fluid branch have indeed been 
extended to temperatures 
below $T_{MCT}$, although the gain is far less spectacular than the one 
claimed in Ref.~\cite{gutierrez}, once thermalization and structural 
stability are more precisely characterized. 

\begin{figure}
\includegraphics[width=\onefig]{fig6a.eps}\hspace{-0.5cm}\llap{\raisebox{2.2cm}{\includegraphics[width=0.8\twofig]{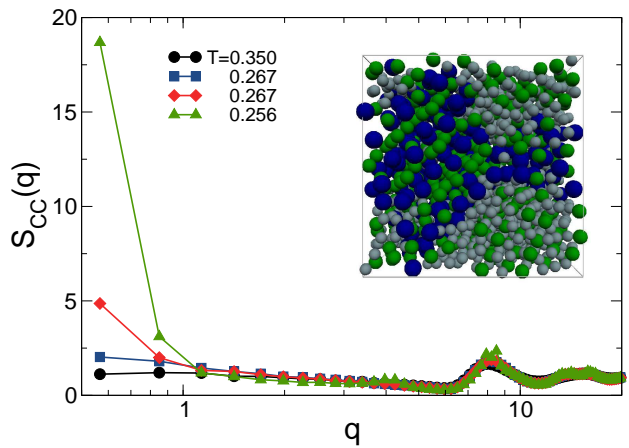}}}
  \caption{Partial structure factor $S_{CC}(q)$ for small particles 
in the ternary mixture of Sec.~\ref{sec:tern}. It is featureless at high enough 
temperatures, $T=0.350$, displays strong composition fluctuations
at low $k$ in the fluid at $T=0.267$, 
that may eventually lead to a demixed state at long times. For $T\leq0.26$,
the system is demixed, as shown for $T=0.256$.
The inset shows a representative snapshot of a demixed 
and partially crystallized system at $T=0.256$.}  
  \label{fig6}
\end{figure}

We studied more carefully the structural properties of the
ternary mixture using partial structure factors, and we show 
some representative results in Fig.~\ref{fig6} for $S_{CC}(k)$
at various temperatures. At high temperatures, $T > 0.30$,
we find that the structure factor resembles the one found for 
ordinary glass-forming models, with a strong first diffraction 
peak corresponding to the inter-particle distance and a featureless
plateau at lower wavevectors. In the low-temperature 
regime, where swap Monte Carlo is mandatory to achieve
thermalization, $0.26 < T \leq 0.29 \approx T_{MCT}$, we find 
that $S_{CC}(k)$ increases more strongly as 
$k$ decreases towards 0,
which suggests that composition fluctuations are 
already quite strong in this regime. 
Even for these state points, which we deemed ``stable'' based on our 
stability criterion, longer simulations show 
that these fluctuations can trigger a demixing 
in the system, as illustrated for $T=0.267$ in Fig.~\ref{fig6}.
We have obtained similarly demixed configurations for temperatures 
up to $T=0.27$, showing that stability is a real issue in this model. 
Finally, for $T \leq 0.26$, the system always demixes within our simulation time, 
which produces a strong low-$k$ peak in the structure factor, see Fig.~\ref{fig6}.
When the particles are segregated, they then easily crystallize 
and we obtain configurations such as the ones shown in the inset 
of Fig.~\ref{fig6}. 
We conclude that maintaining this system in metastable 
fluid state at low temperatures, $T \leq 0.26 \approx 0.9 T_{MCT}$, 
is actually very 
difficult because (i) thermalization becomes 
prohibitively difficult and (ii) simulations longer than the thermalization
time in this regime produce demixed and ordered configurations. 

Finally, to assess more quantitatively 
the gain in efficiency obtained for this ternary mixture, 
we have fitted our dynamic relaxation data from standard 
Monte Carlo simulations to various fitting formula commonly used 
in the glass literature. Using both a Vogel-Fulcher fit, 
Eq.~\eqref{eq:VFT}, which 
presumably overestimates the data at low $T$, 
and a parabolic singularity-free formula, Eq.~\eqref{eq:PARAB}, we 
consistently obtain that the relaxation time 
at $T=0.26$ is about $\tau_\alpha \approx 10^9$. This is 
two orders of magnitude slower than the 
lowest temperature simulated without the swap moves, see Fig.~\ref{fig5}. 
At this low temperature $T=0.26$, the relaxation time using 
swap is $\tau_\alpha \approx 5 \cdot 10^5$ and so the 
thermalization speedup due to swap moves is about 
three orders of magnitude.

Therefore, we confirm that the ternary mixture 
introduced in Ref.~\cite{gutierrez} can be equilibrated at temperatures below the mode-coupling crossover
and we have estimated that this corresponds to an extension of the accessible 
dynamic regime of about two decades compared to standard simulations. This achievement thus 
competes favorably with the other computational approaches described in the 
introduction, but it still does not allow for an exploration 
of glass physics much closer to the experimental glass transition, 
which we shall achieve below for continuous polydispersity.  

\subsection{Five-component mixtures}
\label{sec:pent}

Because ternary mixtures offered only limited success, we tried
to improve both swap move acceptance and the metastability of the 
liquid phase by performing an exploratory study using two different
five-component mixtures. We again adjusted the concentration 
of the various species so that each component roughly occupies
the same volume, and we chose the size ratio between the
different families to be small enough that swap moves are accepted 
with a reasonable acceptance rate.

We studied two systems 
with diameters roughly linearly spaced between $\sigma_{min} = 0.847$
and $\sigma_{\max} = 1.333$ for a polydispersity  $\delta=16\%$,
and between $\sigma_{min} = 0.826$
and $\sigma_{\max} = 1.771$ for a polydispersity  $\delta=23\%$,
Thanks to the reduced size ratio between individual components, the 
swap acceptance rate increased considerably as compared to the binary and 
ternary mixtures studied above, and ranges between 
$a \approx$~10\% and $a \approx$~20\% depending on temperature.
However, both models displayed a strong tendency to demix during swap 
Monte Carlo simulations and it proved impossible to equilibrate these 
systems well below $T_{MCT}$ following the criteria described above.

We have clearly not exhausted all possible discrete models
of glass-formers, as the parameter space becomes very large when the 
number of components increases. It is possible that some 
parameter combination provides both a rapid thermalization 
and a good glass-forming ability, and more work would be needed 
to explore this hypothesis in a more systematic manner, 
as done for instance in the context of simplified models of 
bulk metallic glasses~\cite{zhang2015}.

\section{Continuously polydisperse systems}
\label{sec:poly}

\subsection{Why continuous polydispersity?}

The difficulties highlighted in the previous sections arise from the interplay of several competing effects. Reducing the diameter difference between species improves the acceptance of swap moves, and thus accelerates thermalization, but the resulting reduced polydispersity makes the system prone to crystallization. Additionally, we found that 
simple multi-component mixtures show an important tendency to demixing at low temperature.

To tackle these issues at once, we considered a class of models characterized by a continuous particle size distribution $P(\sigma)$. In such systems, swap moves are more likely to be accepted, because there always exist pairs of particles whose diameters are sufficiently close to one another. 
We found that the succession of a large number of 
successful swaps between pairs with similar diameters
actually facilitates the thermalization of the system. 
Physically, the end result is that the diameter of each particle
performs a kind of random walk in diameter space. An efficient
exploration of the diameter distribution
seems to be the key for efficient
thermalization, as discussed further below in Sec.~\ref{sec:relax}.
 
In addition, by choosing a sufficiently high degree of 
polydispersity it may be possible to stabilize the liquid against 
crystallization and fractionation.
Therefore, well-chosen continuous particle size distributions 
seem able to solve all problems encountered in Sec.~\ref{sec:mix} above 
for discrete mixtures at once.

In this section, we study models 
in which particles interact via Eq.~(\ref{eq:pot}), 
with the cutoff function $F(r_{ij})$ given by Eq.~(\ref{eq:cut}) and a 
cut-off distance $r_{cut}=1.25\sigma_{ij}$. We simulate $N=1500$ 
particles at $\rho=1$. 
We also fix the particle size distribution to be of the 
form given by Eq.~\eqref{eq:pcon} and vary parameters such as the 
pair potential and its additivity. This particle size 
distribution is controlled by a unique parameter, the size
ratio $\sigma_{max} / \sigma_{min}$ or, equivalently, the 
size polydispersity $\delta$. Using insight from 
preliminary studies on hard sphere systems~\cite{beyond}, we fix 
$\sigma_{max} / \sigma_{min} = 2.219$ which implies $\delta \approx 23 \%$. 
This observation is in qualitative agreement with the earlier results of Fernandez {\it et al.}~\cite{Fernandez_Mayor_Verrocchio_2007}, 
who simulated repulsive spheres with a flat size dispersion and found 
an optimal stability for polydispersities in the interval $20\%-30\%$.
In these models, the acceptance of swap moves is typically very high, 
$a \sim 20\%$-$30\%$, and does not change dramatically with temperature. 

\subsection{Influence of the particle softness}
\label{sec:soft}

Our first analysis of models with continuous polydispersity concerns the role of the particle softness.
Previous studies have found that softness can have a non-trivial impact on glass properties, such as fragility~\cite{DeMichele2004} and glass-forming ability~\cite{ohern2015}. 
Here we simulated polydisperse soft particles varying the softness exponent 
$n$ using the values $n=8$, $12$, $18$, $24$. We focus on a continuously polydisperse model with additive interactions, $\epsilon=0$. 
In the limit where $n \to \infty$, the model becomes essentially
equivalent to the hard sphere model studied in Ref.~\cite{beyond},
which displays excellent stability and efficient thermalization. 
Therefore, the present family of models appears as the natural 
extension of these hard sphere results to soft potentials. 

\begin{figure}
  \includegraphics[width=\onefig]{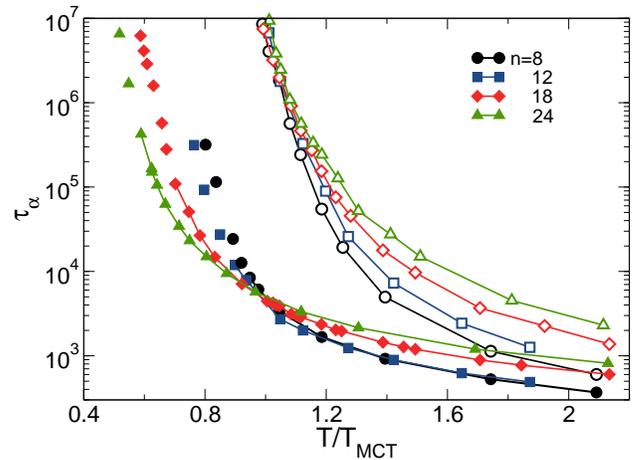}
\caption{Relaxation times for continuously polydisperse systems 
of repulsive soft spheres with different softness 
exponents $n=8$, $12$, $18$ and $24$. Temperatures are scaled by $T_{MCT}$ 
to allow direct comparison between models, with
$T_{MCT}= 0.143$, $0.267$, $0.468$, $0.662$, respectively.
Open symbols represent the standard Monte Carlo dynamics, closed symbols 
the swap algorithm, for which unconnected symbols 
represent structurally unstable 
state points where only a rough estimate of $\tau_\alpha$ is obtained 
in short simulations. A larger $n$ yields better efficiency and 
better structural stability.}
\label{fig7}
\end{figure}

In Fig.~\ref{fig7} we show the structural relaxation times as a function of the temperature for different softness exponents obtained from both standard and 
swap Monte Carlo simulations. 
Temperatures are scaled by the corresponding mode-coupling 
temperature $T_{MCT}$ of each model, so that different systems can be represented in the same graph. In all these systems swap moves speed up thermalization significantly. As the exponent $n$ increases, i.e. as repulsive forces get steeper, 
it becomes possible to equilibrate the system at increasingly 
lower temperatures relative to $T_{MCT}$. 


In addition, as for discrete mixtures, 
the models presented in this section also 
have a tendency to demix at low temperatures. We have again carefully checked the low-$k$ behavior of the partial structure factors that informs 
us about the presence of large composition fluctuations. 
Despite the improved stability, we found that the system may still 
demix at low enough temperatures during the swap Monte Carlo simulations.
For these unstable state points, we again use shorter simulations 
to obtain a rough estimate of the relaxation time, and we show these data 
using disconnected symbols in Fig.~\ref{fig7}. Regarding the structural 
stability of the models, there is again a clear trend with
the softness exponent $n$. We find that softer systems are more prone 
to structural instability than harder ones.
We emphasize that this result is most likely system-specific, since the opposite trend was found in other models of glass-formers~\cite{ohern2015,Douglass_Hudson_Harrowell_2016}.

Combining the effect of a more efficient thermalization and a better 
stability of the fluid state, we conclude that models with larger $n$ values
represent the best choice of parameter within the present family, the system
with $n=24$ being stable and efficiently thermalized down 
to $T \approx 0.6 T_{MCT}$, see Fig.~\ref{fig7}.
The trend that we find suggests that a system of polydisperse hard spheres
with $n=\infty$, such as the one recently simulated in Ref.~\cite{beyond}, 
might actually prove the best glass-former in this class of systems
with continuous polydispersity, repulsive interactions, and additive 
interactions.

\subsection{Non-additive interactions}
\label{sec:nadd}

To suppress the tendency to demixing while preserving 
a continuous form of polydispersity, 
we have introduced non-additivity in the potential by modifying 
the sum rule for particle diameters, as described in Eq.~(\ref{eq:sigNA}). 
Non-additive interactions are known to stabilize the liquid in metallic alloys~\cite{ohern2015,Kob1994}. Moreover, this effect has been demonstrated explicitly in non-additive hard-spheres~\cite{Gazzillo1989,Gazzillo1990}.
Physically, choosing $\epsilon > 0$ frustrates phase separation, 
because particles of different diameters can now stay closer to one another 
than in the additive case with $\epsilon=0$. 
To study the effect of non-additivity, we set the softness 
exponent to $n=12$ and we vary $\epsilon$ using $\epsilon= 0.0,$
$0.1$, $0.2$ and $0.3$. Note that some results for the case $\epsilon=0.2$ 
have been already presented in Sec.~\ref{sec:sim} above, when discussing 
the details of the swap algorithm and the thermalization checks. 

\begin{figure}
  \includegraphics[width=\onefig]{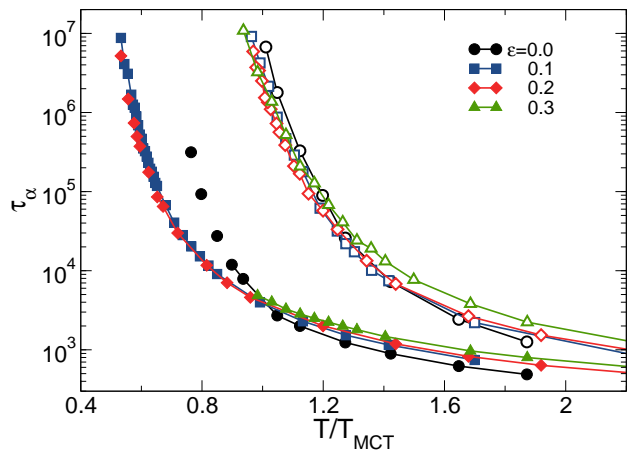}
  \caption{Relaxation times for systems with continuous polydispersity,
$n=12$ and different non-additivity parameter $\epsilon$. 
Temperatures are scaled by $T_{MCT}$ 
to allow direct comparison between models, with
$T_{MCT}= 0.267$, $0.176$, $0.104$, and $0.0534$, respectively.
Open symbols represent the standard Monte Carlo dynamics, closed symbols 
the swap algorithm, for which unconnected symbols 
represent structurally unstable 
state points where only a rough estimate of $\tau_\alpha$ is obtained 
in short simulations.
A well-chosen amount of non-additivity,
$\epsilon \approx 0.1$-$0.2$, considerably
improves the efficiency of thermalization and the structural stability.}
\label{fig8}
\end{figure}

Figure~\ref{fig8} shows the structural relaxation times obtained by varying the degree of non-additivity. As in the previous section, we scale the temperatures by the measured $T_{MCT}$ for each model. 
For comparison, we also redraw the results for the additive system 
($\epsilon=0$) with $n=12$. 
Regarding thermalization efficiency, we find that non-additivity improves
the performance of the swap algorithm, as lower temperatures 
relative to $T_{MCT}$ can be studied when $\epsilon > 0$, but it
is not easy to provide a detailed physical understanding of this result.

Analysis of the low-$k$ behavior of the partial structure factors $S_{\alpha\alpha}(k)$ shows that phase separation is strongly suppressed 
by the non-additivity for both $\epsilon=0.1$ and $0.2$. At low temperature, however, the system can crystallize, as detected from drops in the time series of the energy and from the appearance of well defined peaks in the 
structure factor, and we could frequently observe these 
crystallization events for $\epsilon=0.2$ and $0.3$. 
For $\epsilon=0.1$, the liquid remained stable down to the lowest 
temperature we could equilibrate with the swap Monte Carlo.
This is in fact the only model among the eleven ones studied 
in this paper that perfectly fulfilled our equilibration and stability criteria 
throughout the entire accessible temperature range.

The existence of an optimal non-additivity parameter $\epsilon$
for glass-forming ability presumably results from a competition
between demixing, which takes place when $\epsilon=0$, and 
crystallization, for large $\epsilon$.
We suggest that the existence of these two distinct paths 
to ordering actually compete for $\epsilon=0.1$-$0.2$, which 
results in an enhanced frustration and thus a higher structural stability.
Similar physical arguments have been proposed in Ref.~\cite{zhang2015}
to explain glass-forming ability of simple non-additive mixtures.

Whereas the additive model with $n=12$ was not the best choice 
of softness in the previous section, we find that including non-additivity 
significantly improves both the efficiency of the swap algorithm and 
the structural stability of this model. We suggest that models 
with larger $n$ (including hard spheres) with non-additive 
diameters would be even better choices. 

It is interesting
to contrast the results for the $n=12$ soft repulsion 
using a continuous size distribution and a non-additivity $\epsilon=0.1$
to the results obtained for the binary mixture in Sec.~\ref{sec:bin}
for the same pair potential. The conclusion is that the 
thermalization and stability limits have been 
decreased from $T \approx T_{MCT}$ for the binary mixture down to 
$T \approx 0.5 T_{MCT}$ for the present system. This
represents a major methodological improvement that we try to quantify
in terms of timescales in the next subsection.

\subsection{Experimental timescales are matched
by simulations}

In the previous section, we showed that by optimizing the additivity 
and the form of the pair potential, temperatures as 
low as $T \approx 0.5 T_{MCT}$ could be thermalized using swap 
Monte Carlo in the metastable fluid 
state. Because ordinary simulations stop near $T \approx T_{MCT}$, 
one may wonder how large is the corresponding gain in terms of structural relaxation times.
It is relatively easy to answer this 
question when the improvement is modest, but
this becomes a delicate task in our case, as thermalization is achieved 
in a temperature regime where the standard dynamics is completely frozen in our observation window and
where equilibrium timescales can only be obtained by extrapolation. 
Extrapolating timescales down to the lowest temperatures where the swap 
Monte Carlo can thermalize 
may depend sensitively of the fitting procedure and it therefore requires some care. 

We have devised a robust strategy which answers
for each model the following question: Is the thermalization speedup due to the 
swap Monte Carlo algorithm large enough to fill the eight-decade gap 
between ordinary simulations and experiments?
To answer this question for a given model,
we employ ordinary Monte Carlo simulations to access a range of relaxation time 
up to $\tau_\alpha / \tau_0 \approx 10^4$, where $\tau_0$ represents
the value of $\tau_\alpha$ at the onset of glassy dynamics. 
In experiments~\cite{Rossler2005}, the glass temperature $T_g$ corresponds to 
the value $\tau_\alpha / \tau_0 \approx 10^{12}$, which we will take 
as our practical definition of $T_g$.
Using the various functional forms described in Sec.~\ref{sec:phys}, we realized that estimating 
the location of $T_g$ from numerical measurements of $\tau_\alpha$ 
is actually possible with modest uncertainty.
More precisely, for a given model we use all three functional forms in 
Eqs.~(\ref{eq:VFT}, \ref{eq:PARAB}, \ref{eq:ARRH}) to estimate the location
of $T_g$ from the definition $\tau_\alpha (T_g) / \tau_0 = 10^{12}$.  
Despite the qualitative 
differences between these functional forms, the range 
of $T_g$ values is reasonably small, typically $\Delta T_g / T_g = (T_g^\textrm{VFT}-T_g^\textrm{Arrhenius}) / (2T_g^\textrm{parabolic}) \approx
12 \%$, with minor variations from one model to the other. 
Because the VFT law tends to overestimate the increase $\tau_\alpha$ and the Arrhenius law underestimates it, these two forms respectively provide 
an upper and a lower bound to the real location of $T_g$, while estimates from the parabolic law usually fall between those two bounds.
Elmatad \textit{et al.}~\cite{Elmatad2009} have shown that the parabolic law accounts for the variation of relaxation times of glass-forming liquids over a broad range of temperatures, ranging from the onset down to the laboratory glass transition.
We therefore expect our parabolic extrapolation to 
provide a reasonable determination of $T_g$, and the other two temperatures 
to provide a solid estimate of the interval of confidence. 

\begin{figure}
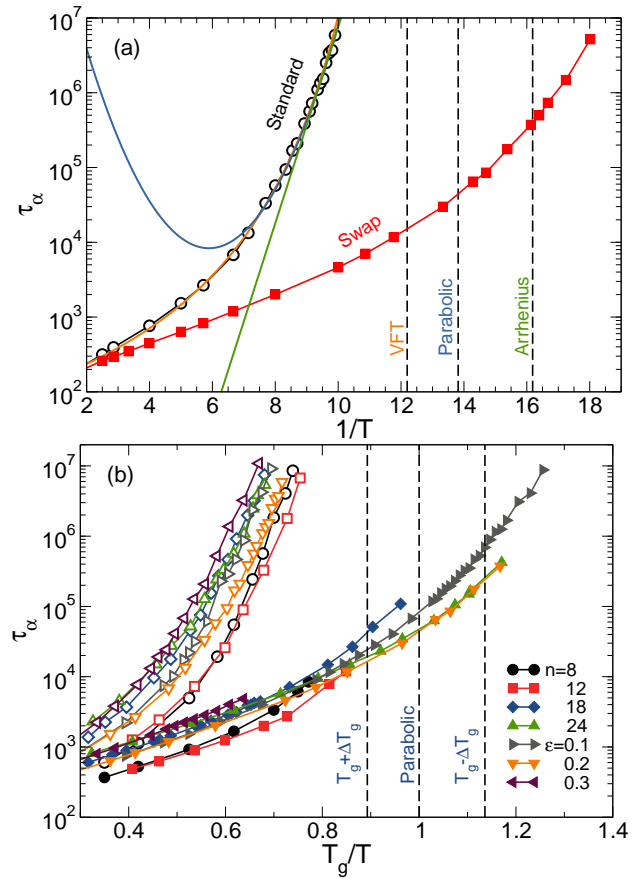

  \includegraphics[width=0.98\onefig]{fig9a.eps}
  \includegraphics[width=\onefig,clip]{fig9b.eps}
  \caption{(a) Relaxation times for the non-additive model with 
$n=12$ and $\epsilon=0.2$ for standard and swap Monte Carlo dynamics.
The standard dynamics is fitted with the VFT, parabolic and Arrhenius laws, 
as shown with lines, which are used to estimate the location 
of the experimental glass temperature $T_g$, as shown with 
vertical dashed lines. For this system, the swap dynamics is able to provide 
stable and thermalized configurations at temperature below $T_g$. 
(b) Relaxation times obtained from standard (open symbols) and
swap (filled symbols) dynamics for various size 
polydisperse models of various softness ($n$) and non-additivity ($\epsilon$) 
are shown in an Arrhenius form with rescaled temperature $T_g/T$, where 
$T_g$ is estimated as in (a). For all models the thermalization 
speedup near $T_g$ is of about 10 orders of magnitude, some models
being structurally stable down to temperatures below $T_g$.}
\label{fig:ceiling}
\end{figure}

We illustrate this procedure  in Fig.~\ref{fig:ceiling}(a) where we show 
the temperature evolution of $\tau_\alpha(T)$ 
from standard Monte Carlo dynamics for 
the non-additive model studied in Sec.~\ref{sec:nadd}, using 
an Arrhenius representation.
We fit these Monte Carlo dynamic data and estimate 
three different locations for $T_g$, which delimit the 
range of possible values for the location of $T_g$, highlighted
with the vertical dashed lines. 
We then show the evolution of the relaxation time 
for metastable fluid states when swap Monte Carlo moves are used
in the same Arrhenius representation. 
We find that the swap relaxation time remains modest in the vicinity of 
$T_g$, $\tau_\alpha / \tau_0 \approx 10^2-10^3$ and that this 
particular system can be thermalized and maintained metastable
at temperatures even below 
the experimental glass temperature $T_g$. This finding  
has several important consequences. 

\begin{itemize}

\item
The speedup 
of the thermalization at $T_g$ is of about 10 ($\pm 1$) orders of magnitude, 
implying that computer simulations can now 
comfortably study that temperature regime
in thermal equilibrium.

\item
The maximal speedup obtained with the swap Monte Carlo is in fact  
much larger than these 10 orders of magnitude,
because temperatures lower than $T_g$ can be thermalized,
but estimating that gain becomes very sensitive to the 
chosen extrapolation. 
 
\item
For selected models we can now access a temperature regime that even experiments cannot reach, thus opening
a novel observational window on the physics of glasses in a regime 
that has never been probed before, either experimentally or numerically.

\end{itemize}

To quantify the performance of the swap algorithm, 
we estimate the range of $T_g$ values for each model studied in this 
section, and use these fitted values to construct an Angell plot, 
representing the logarithm of the relaxation times as a function 
of the scaled inverse temperature, $T_g/T$, see 
Fig.~\ref{fig:ceiling}(b). In practice, we use 
the value given by the parabolic which falls in the middle of the fitted 
range and show the corresponding uncertainty, estimated from 
VFT and Arrhenius fits, with the vertical 
dashed lines in Fig.~\ref{fig:ceiling}(b).
Standard Monte Carlo simulations typically stop near 
$T/T_g \approx 1.3$-1.5, in the vicinity of the mode-coupling crossover.
For most models, the swap algorithm performs so well that the 
thermalization time at $T_g$ remains modest, 
$\tau_\alpha/\tau_0 \approx 10^2$. This corresponds to a thermalization
speedup at $T_g$ of 10 orders of magnitude.
As mentioned before, models with soft potentials and 
additive interactions are prone to structural instability, and some
of them are not stable down to $T=T_g$. However for several models, 
we find that thermalization and fluid metastability can be maintained
below $T_g$. 

The discovery of such glass-forming models associated to an efficient
algorithm to thermalize them represents the main achievement of our work.

\section{Microscopic insights into the swap dynamics}

\label{sec:relax}

\subsection{Dynamics of particle diameters}

The previous sections demonstrated that swap Monte Carlo moves can enhance 
thermalization by several orders of magnitude, the effect being most 
spectacular in continuously polydisperse systems, 
for which swap moves have a very high acceptance rate.
In this section we shed some light on the microscopic
mechanisms that are responsible for this acceleration.
We carry out this analysis for a non-additive polydisperse 
model with $\eps=0.2$ introduced in Sec.~\ref{sec:nadd}.

It has been previously suggested that the swap moves increase 
the particle mobility because they allow the particles 
to escape the cage formed by their 
neighbors~\cite{Grigera_Parisi_2001, gutierrez} after a non-local
swap move. This view seems correct when one considers that 
particles exchange their positions, because a caged particle 
indeed appears to jump instantaneously to a novel position. However, the 
swapped particle is actually replaced by another particle which is then
occupying the caged position itself, and it jumps to a position  
where another particle was caged too. Therefore, it is not 
clear that the cage is affected at all after a swap move, and 
this simple explanation cannot explain the speedup of the dynamics. 

This conclusion is more easily 
grasped when one considers, as we do, that particles simply exchange their
diameters during a swap move, without changing position. In that case, 
the diameter of each caged particle slowly fluctuates in time.
For continuous polydisperse systems, these time fluctuations 
take the form of a random walk in diameter space. Therefore, we
conclude that it is rather the slow wandering of the
diameter of each particle that allows the system to relax more efficiently 
towards equilibrium. A naive physical explanation would be that 
a caged particle with a large diameter could start diffusing 
by shrinking its radius, thus being able to squeeze and escape through 
a small channel. We shall now demonstrate that the physics is 
actually more complicated and more collective than this naive image.

\begin{figure}
  \includegraphics[width=\onefig]{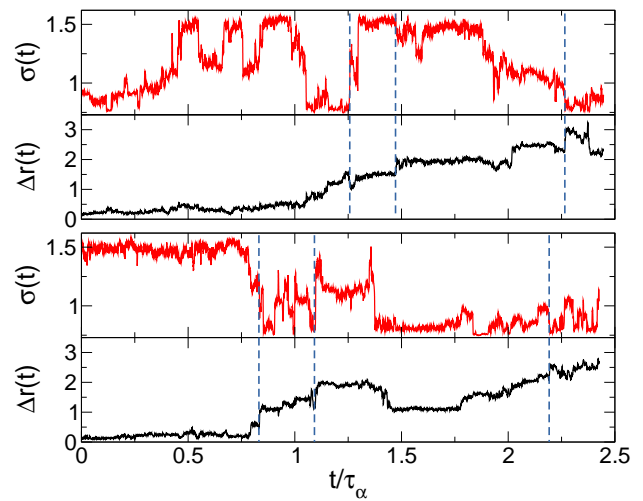}
  \caption{Time-series of individual 
displacement $\Delta r(t)$ and diameter value $\sigma(t)$ for 
two tagged particles in the non-additive model of Sec.~\ref{sec:nadd}
with $\epsilon=0.2$ at $T=0.0555$.
Intermittent diffusion 
in real and diameter space is observed, with strong correlations between 
$\Delta r(t)$ and $\sigma(t)$ highlighted with dashed lines, but we 
also observe many events in one observable that have no counterpart 
in the other indicating that the correlation between the two observables 
is non-local.}
\label{fig9}
\end{figure}

To see this, let us start with some qualitative observations on the time evolution of the diameter of a tagged particle. We show two typical time series of $\sigma_i(t)$ for two different particles in Fig.~\ref{fig9}. We rescale the time 
axis by $\tau_\alpha$ to better appreciate how much the particle 
diameter changes over a relaxation time of the system. 
On short time scales the diameter fluctuates around its initial value, while at longer times it changes and eventually visits all values allowed by the particle size distribution. At low temperature these relaxation events occur suddenly and appear as jumps. Overall, this behavior strongly resembles the typical features of glassy dynamics known from real space analysis of single particle displacements, mimicking very much the cage effect and hopping motion. To reinforce this analogy, we show the time series of the particle displacements for the same two tagged particles in Fig.~\ref{fig9}. As expected, they display periods of immobility separated by rapid jumps.

The comparison of the two panels reveals that some of the sudden jumps in 
diameter space occur at similar times as the sudden jumps of the particle in 
real space which indicates that diameter dynamics can trigger 
diffusion.
However, we can also detect jumps occurring in real space without clear 
counterparts in diameter space, and vice versa. 
These observations suggest that changing the diameter of a single particle is 
not necessarily enough to trigger a rearrangement, and also that 
changes in the neighborhood of one particle may be enough to trigger a 
displacement.
Overall, the physical picture is that relaxation in these
supercooled states is a collective process 
and the efficient thermalization with swap 
cannot be explained on the basis of a simple single-particle argument.

\begin{figure}
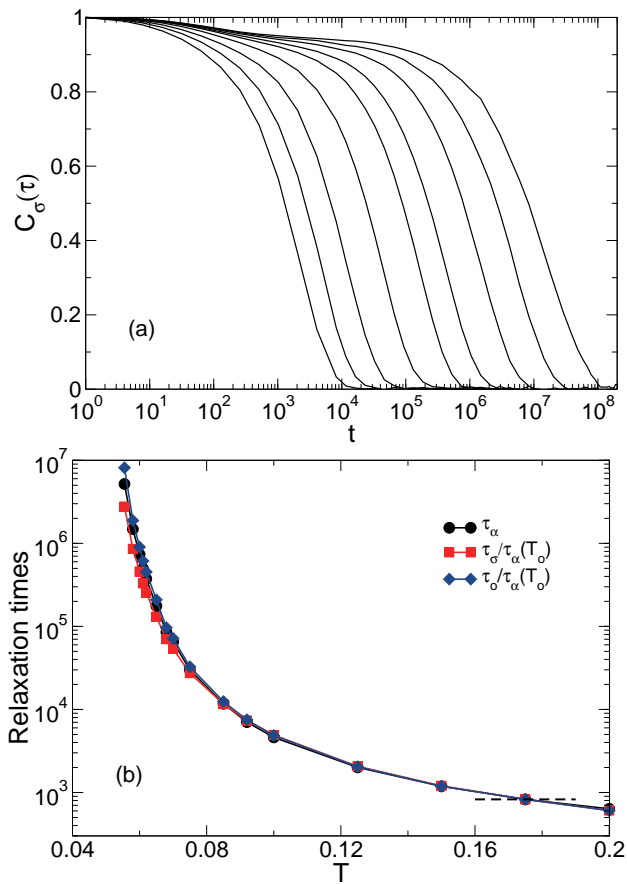

  \centering
  \includegraphics[width=0.98\onefig]{fig11a.eps}
  \includegraphics[width=\onefig,clip]{fig11b.eps}
  \caption{Non-additive model of Sec.~\ref{sec:nadd}
with $\epsilon=0.2$.
(a) Time auto-correlation of particle diameters $C_\sigma (t)$ 
measured during the swap dynamics for temperatures as in Fig.~\ref{fig2}.
(b) Relaxation times $\tau_\alpha$, $\tau_\sigma$ and $\tau_o$ 
as a function of temperature, with $\tau_\sigma$ and $\tau_o$ rescaled 
to coincide with $\tau_\alpha$ at $T=0.175$ (shown with horizontal bar). 
The three timescales 
obviously have the same temperature dependence.}
  \label{fig11}
\end{figure}

To quantify the correlation between diameter and position dynamics, 
we define a time correlation associated to the time evolution 
of the diameters. We define the following auto-correlation function
for diameter fluctuations:
\begin{equation}
\label{eq:corrS}
C_\sigma(t) =\left< c_\sigma(t) \right> = \left<\frac{\sum_i\delta\sigma_i(t)\delta\sigma_i(0)}{\sum_i \delta\sigma_i^2(0)}\right>,
\end{equation}
where $\delta \sigma_i(t)= \sigma_i(t) - \langle \sigma \rangle$. 
This function is normalized so that it evolves from unity when $t=0$, 
to zero at large times when diameters become completely uncorrelated from their initial values.
The temperature variation of this function is shown in Fig.\ref{fig11}(a). By lowering the temperature $C_\sigma(t)$ develops a plateau after an initial short time decay and eventually decorrelates at longer times. This confirms the previous qualitative observations that the diameters remain ``caged'' around some initial value before complete decorrelation.

Let us define the diameter decorrelation times $\tau_\sigma$ as the value of time such that $C_\sigma(\tau_\sigma)=e^{-1}$. In Fig.~\ref{fig11}(b), we compare
the temperature evolution of three different relaxation times 
for single-particle motion, $\tau_\alpha$, for single-particle diameter, 
$\tau_\sigma$, and for collective density fluctuations, $\tau_o$. 
We absorb the observable dependence of these three timescales by rescaling them
at a single temperature where the relaxation is fast, namely $T=0.175$. 
The striking result of these measurements is that single particle displacements,
density fluctuations and diameter fluctuations all relax on the same timescale.
Because diffusion is fully arrested when the diameters do not fluctuate
we conclude that it is the efficient dynamics in diameter space which 
drives the structural relaxation in position space, and therefore, the 
efficient thermalization of the system.

A further intriguing observation about the role of 
diameter fluctuations stems from the data shown in Fig.~\ref{fig2},
where time correlation functions for standard and swap dynamics 
are compared at the same very low temperature, where even the swap dynamics
is very slow. In that case one observes that 
the plateau height related to short-time vibrational motion 
is different in the two dynamics, 
the amplitude of these vibrations being much larger when
the swap dynamics is used. This observation implies that 
at short times the small fluctuations in particle diameters
act as an additional degree of freedom that allows each particle
to perform back and forth caged motion over a typical distance
that is larger than in the standard dynamics. These larger in-cage fluctuations
suggest a possible ``softening'' of local cages, which 
seems to correlate well with an acceleration of the dynamics. 
Such a correlation between short-time motion and structural 
relaxation is often discussed in the context of glass-forming
models~\cite{Buchenau,Scopigno2003,Dyre_2006,Hansen_Frick_Hecksher_Dyre_Niss_2017}, and it would be interesting to study it further 
in the present context. 

\subsection{Spatially heterogeneous dynamics}

The correspondence between the timescales for diameter and position dynamics, 
accompanied by a lack of strong correlation at the single particle level, 
suggests that the physics of diffusion in real and diameter space 
is cooperative in nature. 
For instance, diffusive events could be happening more easily in a 
spatial region where 
the diameter dynamics has been particularly efficient. This hypothesis 
suggests to investigate the existence of spatial correlations of the 
dynamical relaxations. 

\begin{figure}
  \includegraphics[width=\twofig]{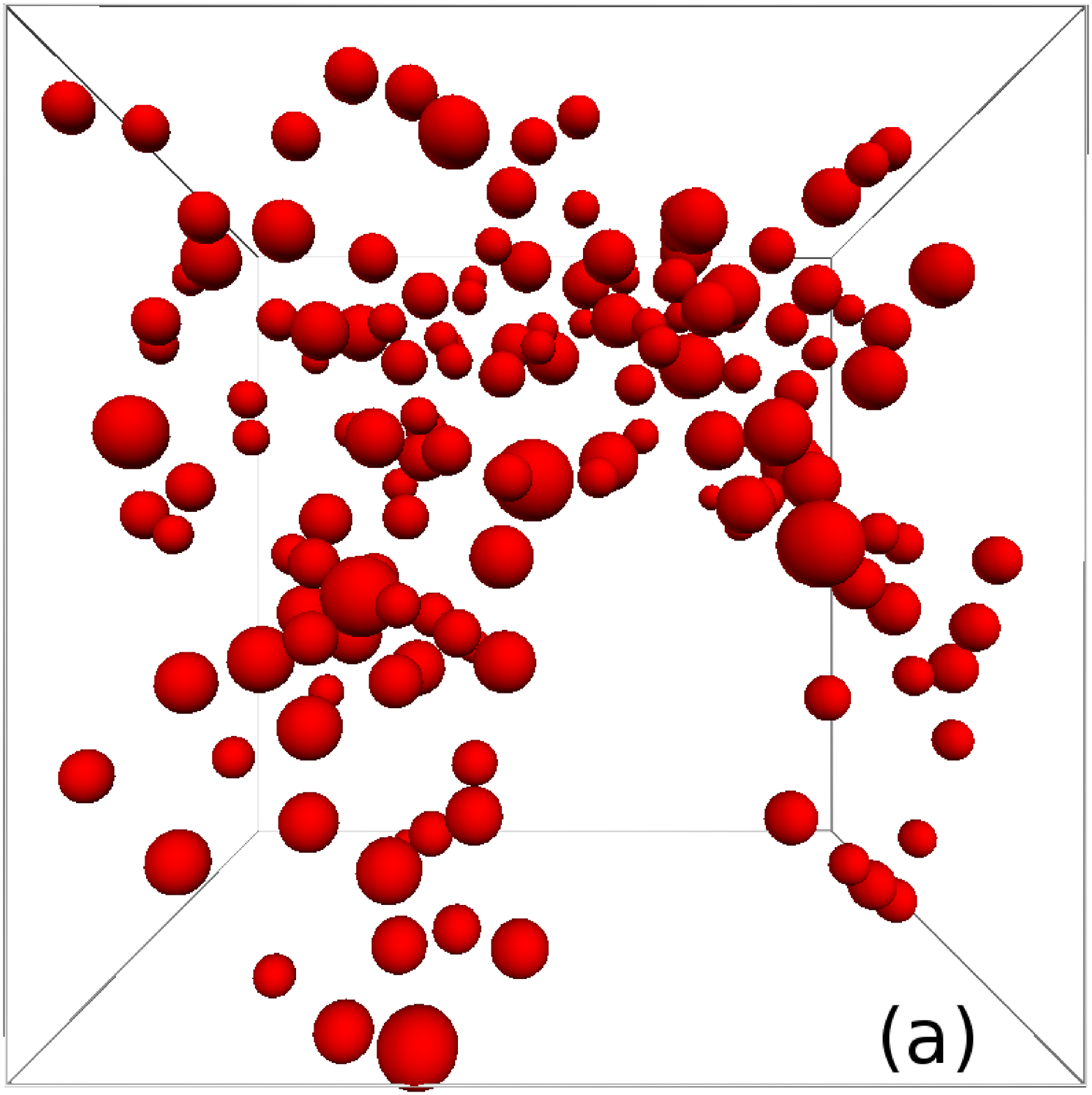}
  \includegraphics[width=\twofig]{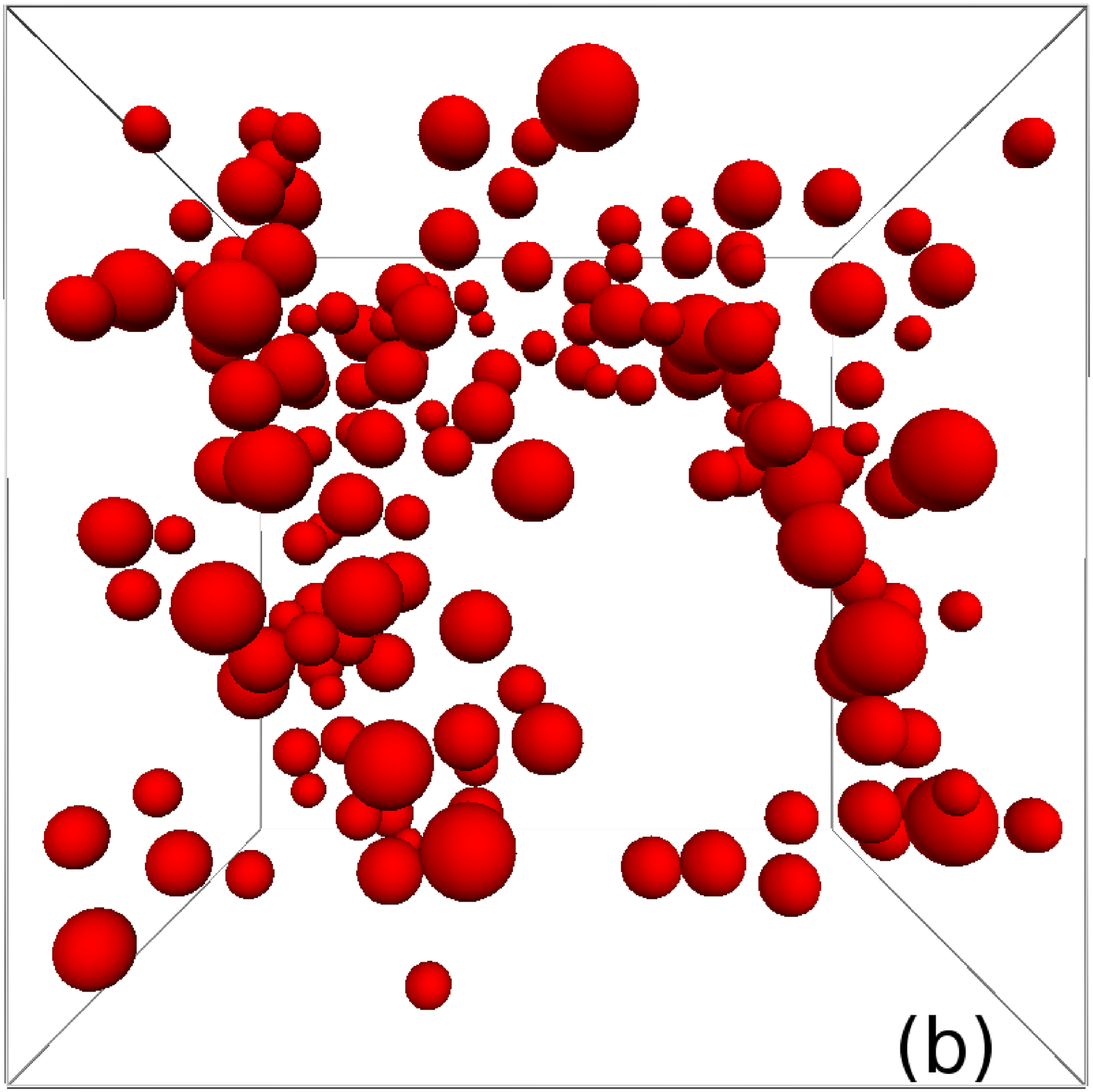}
  \caption{\label{fig10} Snapshots of the 10\% of particles with 
(a) the largest displacements (b) the largest diameter change,
computed between times $t=0$ and $t=\tau_\alpha/2$ for 
$T=0.0555$. There is a clear correlation between the spatial regions where 
dynamics in real and diameter spaces are fast, but the correlation is 
weak at the single-particle level.} 
\end{figure}

To illustrate this point qualitatively, we show in Fig.~\ref{fig10} 
two typical configurations at a temperature $T=0.0555$.
We measure the dynamics between an arbitrary initial time
$t=0$ and a later 
time $t = \tau_\alpha(T) / 2$. In Fig.~\ref{fig10}(a) we show 
the particles having the 10$\%$ largest displacements in real space 
over this time lag, whereas in Fig.~\ref{fig10}(b)
we show 
the particles having the 10$\%$ largest displacements in diameter space. Particles are drawn using rescaled final diameter values.
We observe a close similarity between regions of faster diffusing 
particles and regions of particles with large diameter changes, 
but we also recognize that the correlation does not hold 
at the particle level. Thus, we conclude that diameter changes and structural relaxation may affect each other in a non-local fashion. 

\begin{figure}
    \includegraphics[width=\onefig]{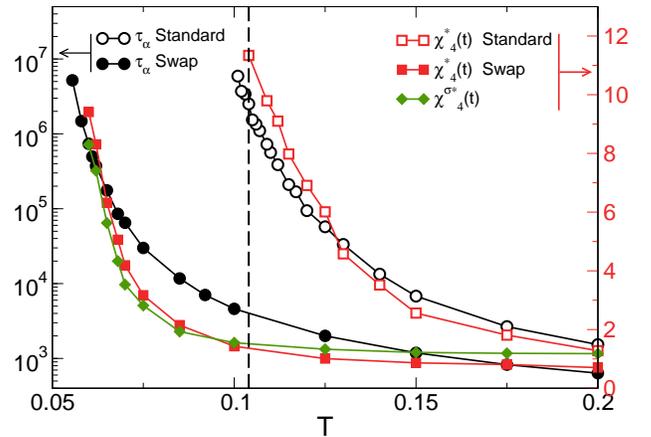}
    \caption{Temperature evolution of dynamic susceptibilities (right axis)
and relaxation times (left axis), the vertical dashed line is 
$T_{MCT} = 0.104$.
Whereas $\chi^d_4(t)$ grows together with $\tau_\alpha$
in standard simulations, $\chi^d_4(t)$ and $\chi_4^\sigma(t)$
behave similarly and have a very different temperature dependence which 
mirrors instead the evolution of the swap relaxation time.}
\label{fig13}
\end{figure}

The spatial correlations of diameter fluctuations can be characterized using the multi-point functions introduced to study cooperative motion in supercooled liquids~\cite{toninelli}.
The generic expression of the dynamical susceptibility 
related to a time-dependent observable  $O(t)$ is
\begin{equation}
\chi^O_4(t)= N\left[ \langle O^2(t) \rangle - \langle O(t) \rangle^2 \right].
\end{equation}
It quantifies the extent of spatial correlations associated to the local observable $O$ over a time scale $t$~\cite{toninelli}.
Here we measure dynamic susceptibilities associated both to 
the self-part of density fluctuations, $\chi^d_4(t)$ with $O(t)= f_s(k,t)$ 
[see Eq.~\eqref{eq:self}], and to diameter fluctuations, $\chi^\sigma_4(t)$ with $O(t)= c_\sigma(t)$ [see Eq.~\eqref{eq:corrS}]. 
These functions provide information on the spatially heterogeneous dynamics 
of particle displacements and of diameter changes, and they typically display 
a peak around the timescales $\tau_\alpha$ and $\tau_\sigma$, respectively. 
In Fig.~\ref{fig13} we report the height of these two peaks, 
$\chi^{d*}_4$ and $\chi^{\sigma*}_4$, as a function of the temperature. 
In addition we also measure and report the behavior of $\chi^{d*}_4$ 
for the standard Monte Carlo dynamics.

This figure provides two main pieces of information. 
First, we notice that the temperature dependence of $\chi^{d*}_4$ in standard and swap simulations are very different. The behavior for standard Monte Carlo is as reported before~\cite{Berthier_Biroli_Bouchaud_Kob_Miyazaki_Reichman_2007b}, where
 $\chi^{d*}_4$ increases rapidly from a value 
$\chi^{d*}_4 \approx 1$ when the temperature is decreased below 
the onset $T_o \approx 0.18$ to a value $\chi^{d*}_4 \approx 12$
when approaching $T_{MCT}$, in way that mirrors the evolution of the
relaxation time $\tau_\alpha$, as demonstrated in Fig.~\ref{fig13}.
The traditional interpretation is that 
dynamics becomes spatially correlated over larger length-scales as 
the temperature is lowered.  
A striking observation is that the swap dynamics near $T_{MCT}$ 
displays essentially no spatial dynamic correlations. By construction, 
swap moves can affect the dynamics of the system but not its 
equilibrium static properties. Therefore, we conclude that 
the growth of the spatial correlations detected by $\chi^{d*}_4$ 
for $T > T_{MCT}$ in standard dynamics is mostly of dynamic,
rather than structural, origin. This finding, which agrees qualitatively with 
previous conclusions~\cite{berthierJack,jack2014}, may also explain why the swap algorithm can be
very efficient, because if spatial correlations had a strong 
structural component, then a strong numerical acceleration 
would likely necessitate
the introduction of a more collective algorithm.

The second key information from Fig.~\ref{fig13} is that both quantities 
$\chi^{d*}_4$ and $\chi^{\sigma*}_4$ with swap dynamics 
are quantitatively very close, which confirms that fluctuations 
in real and in diameter spaces are strongly correlated.
Even though the correlation is not strong at the local scale, diameters fluctuations display the same temperature evolution as dynamic heterogeneities.
In addition, both quantities follow the growth of the swap relaxation time, 
the swap algorithm becoming slow at low enough temperatures, at which 
important 
spatial correlations of the diameter dynamics become needed to relax to 
system towards equilibrium. 

\section{Ideas for the future design of 
glass-forming models}
\label{sec:design}

In this section, we build on the detailed level 
of understanding of the swap mechanism reached in the previous 
sections to propose novel directions and ideas to design 
new glass-forming models for which the swap Monte Carlo 
approach could be very efficient.

\subsection{``Hybrid'' models for binary mixtures}
\label{sec:hybrid}

A large number of models studied in the past were based on discrete 
mixtures~\cite{bernu,Kob1994,Wahn}, as studied in Sec.~\ref{sec:mix}. We concluded there 
that swap was not well-suited for binary mixtures because 
a large acceptance rate for the swap seems 
incompatible with a good structural stability. 
We now show that it is possible to construct models which 
have the characteristics of binary mixtures, reasonable 
stability, and can be efficiently simulated using the swap algorithm.

Our idea is to introduce what we call a ``hybrid'' particle 
size distribution, as sketched in Fig.~\ref{fig1}. These
distributions are composed 
of two main peaks which are a good representation of an
$A-B$ binary mixture. 
In order to have a good glass-forming ability, we choose an equal 
concentration of particles in these two peaks, and, more importantly, 
we choose a size ratio which is large enough to avoid 
the crystallization observed otherwise. Because such a large size ratio
implies that $(A,B)$ swap moves are always rejected, we introduce a third specie
in the model, associated to a flat continuous
distribution of particle sizes that connects smoothly the two main peaks of the binary mixture. 
The main idea is that a particle belonging
to one of the two main components can be swapped with particles
belonging to the intermediate 
third specie, and it can then slowly tunnel through to reach 
the other specie. In other words, whereas a direct particle exchange
between $A$ and $B$ 
is unlikely, the addition of the interpolating specie facilitates such 
exchanges which can then happen via a large number of intermediate 
swaps which all have a large acceptance rate.   

In practice, we introduce two species ($A,B$) with flat continuous polydispersity around two average diameter values ($\sigma_A, \sigma_B$) such 
that $\sigma_B/ \sigma_A = 1.6$. 
We add a third specie, $C$, with an average diameter value $\sigma_C=(\sigma_A+\sigma_B)/2$, 
which continuously interpolates between small and large diameters. 
Each specie contains roughly $\frac{1}{3}$ of the particles.
The final size distribution is described by Eq.~(\ref{eq:phyb}) 
with the chosen parameters  $x_A=0.33$, $x_B=0.34$,  $x_C=0.33$, $\sigma_A=0.76$, $\sigma_B=1.23$, $\sigma_C=1.00$, $b_A=0.04$, $b_B=0.04$, $b_C=0.26$.
The two-body potential is given by Eqs.~(\ref{eq:pot}, \ref{eq:cut}) 
with $n=12$ and a cut-off distance of $r_{cut}=1.25\sigma_{ij}$. 
We perform simulations with $N=1000$ at $\rho=1.3$.
With these parameters, the  
polydispersity is $\delta = 20\%$. 

\begin{figure}
  \includegraphics[width=\onefig]{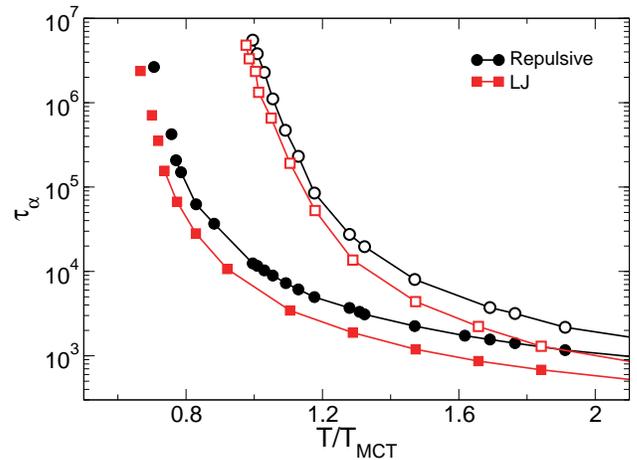}
  \caption{Relaxation times for repulsive and Lennard-Jones potentials
with a hybrid particle size distribution.
Temperatures are scaled by $T_{MCT}$ to allow direct 
comparison between models, with $T_{MCT} = 0.680$ and $0.543$
for repulsive and LJ potentials, respectively.
Open symbols represent the standard
Monte Carlo dynamics, closed symbols the swap algorithm,
for which unconnected symbols represent structurally unstable 
state points where only a rough estimate of $\tau_\alpha$ is obtained
in short simulations.}
  \label{fig14}
\end{figure}

Results for the relaxation times are presented in Fig.~\ref{fig14}. As usual, we use disconnected points to represent unstable state points for which short simulations are used to estimate $\tau_\alpha$. For this system again, 
we observe that equilibration is easily attainable for temperatures below $T_{MCT}$, and resistance to ordering is ensured down to relatively low 
temperatures, $T \approx 0.75 T_{MCT}$. 
However, at lower temperatures the system again present instabilities 
due to the tendency to phase separation that we observe through low-$k$ 
values of the structure factor. Preliminary results indicate that 
using non-additive interactions will most certainly stabilize the 
system down to even lower temperatures, but the main 
goal of this section has nevertheless been reached. We have indeed designed
a model which is structurally similar to an 
equimolar binary mixture with 
size ratio 1.6, but that can be efficiently studied using the swap algorithm 
down to $T = 0.75 T_{MCT}$ via the introduction of a third, 
intermediate specie.
Using the same fitting procedure as above, we estimate that 
these low temperatures allow us to access a total dynamic range of about 
$\tau_\alpha / \tau_0 \approx 10^8$, 
so that the swap Monte Carlo method already allows the 
exploration of a novel temperature regime corresponding to an increase
in relaxation times of about 3-4 orders of magnitude as compared to standard
Monte Carlo simulations.
It would be interesting to study hybrid variants of discrete mixtures, in which the concentration of specie $C$ is small enough to be considered as a 
small perturbation of the original model. Work in this direction is in progress.

\subsection{Lennard-Jones interactions}

Up to now, we have studied pair potentials describing  
repulsive soft spheres with inverse power law repulsion of various softness. 
However, more realistic pair interactions including attractive 
forces are often used in studies of supercooled liquids.
Perhaps the most studied pair interaction is the Lennard-Jones potential~\cite{Kob1994},
which contains a soft power law repulsion with exponent $n=12$, combined with
a soft power law attraction with exponent $n=6$, see Eq.~(\ref{eq:LJ}).

In this final study, we test whether Lennard-Jones interactions 
can also be efficiently studied using the swap Monte Carlo 
algorithm. To this end, we start from the previous ``hybrid'' model 
studied in Sec.~\ref{sec:hybrid} and include a power law 
attraction. Because the potential is now longer-ranged, we use a
larger cutoff  $r_{cut}=2.5\sigma_{ij}$, and shift the potential by the constant
$c_{LJ}$ to ensure the continuity of the potential at the cut-off. 
All other parameters are equal to the ones employed in the 
hybrid repulsive model above in Sec.~\ref{sec:hybrid}.

We again perform a comparison of the standard and swap dynamics for 
this Lennard-Jones system, and present the results along the ones
of the corresponding repulsive case in Fig.~\ref{fig14}.
We find that including attractive forces modifies very little 
both dynamics, apart from a rescaling of the temperature 
scale: the mode-coupling crossover temperature shifts 
from 0.680 to 0.543 when including attractive forces~\cite{gilles}.
As a result, the above conclusions regarding stability 
and thermalization efficiency directly carry out to this 
Lennard-Jones system. 
Our main conclusion is therefore that
our ``feasibility study'' is successful and that  
glass-forming models with Lennard-Jones interactions
and a binary-like size distribution
can be devised and studied down to very low temperatures
using the swap algorithm.
Such models will most certainly prove useful in future studies
of the glass transition.
 
\section{Perspectives}
\label{sec:discuss}

In this article, we established that a number of 
glass-forming models with various pair interactions,  
particle size distributions and degree of non-additivity,
can be efficiently simulated using a simple
swap Monte Carlo algorithm and remain excellent glass-formers down to very low
temperatures. For some models, we have been able to thermalize 
the metastable fluid down to temperatures that are 
lower than the laboratory glass transition, 
which represents the current experimental limit for molecular liquids. 
Therefore, our paper not only fills the 
eight orders of magnitude gap between ordinary simulations 
and experimental work, but it actually goes beyond state-of-the-art
experiments and demonstrates that 
both static and short-time dynamical properties
can now be studied in computer simulations in a novel temperature regime.
In addition to static quantities, 
by using thermalised configurations obtained with the 
swap method as initial conditions for trajectories generated without swap, 
we believe it is possible to substantially 
extend the dynamic window for structural relaxation, which may shed new 
light on the glassy dynamics as well. 

Our achievements are summarized in Fig.~\ref{fig1}, but throughout the 
article we have suggested several ways in which our approach could 
be extended to devise different or more realistic models
of glass-forming materials. We have also suggested ways in which 
the algorithm itself could be improved and described 
several paths that remain to be explored in future work.
We expect these results to trigger a large research activity 
towards these goals.  

Obtaining thermalized states in simple models of supercooled 
liquids at temperatures comparable to the experimental glass transition
paves the way to a number of novel studies, because 
essentially all simulation work published over the past 30 years could 
be performed again over a previously 
inaccessible temperature regime. Some works 
along these lines have been already published~\cite{beyond,gardner}, and others are currently in progress regarding the thermodynamic properties of deeply supercooled liquids, their local structure,
vibrational and mechanical properties, and the existence of a Gardner 
transition in soft glasses. 

\acknowledgments
We thank 
G. Biroli, 
R. Jack, 
M. Ozawa, 
I. Procaccia,
G. Tarjus, and
M. Wyart for useful exchanges about this work.
We thank R. Gutierrez for providing 
additional information regarding simulations 
performed in Ref.~\cite{gutierrez}.
The research leading to these results has received funding from the 
European Research Council under the European Unions
Seventh Framework Programme (FP7/2007-2013)/ERC
Grant Agreement No. 306845. This work was supported
by a grant from the Simons Foundation (\# 454933, Ludovic Berthier).

\bibliography{draft}

\end{document}